\documentclass[aps,twocolumn,showpacs]{revtex4}
\makeatletter

\newcommand{\Rmnum}[1]{\expandafter\@slowromancap\romannumeral #1@}
\makeatother
\usepackage{graphicx}% Include figure files
\usepackage{dcolumn}% Align table columns on decimal point
\usepackage{bm}% bold math
\usepackage{CJK}
\usepackage{amsfonts}
\usepackage{pifont}
\usepackage{psfrag}
\usepackage{wrapfig}
\usepackage{subfigure}
\usepackage{makeidx}
\usepackage{multirow}
\usepackage{epsf}
\usepackage[colorlinks,linkcolor=red,anchorcolor=red,citecolor=blue,urlcolor=blue]{hyperref}
\usepackage{amsmath}%»¨Ìå×Öĸ
\usepackage{cases}
\usepackage{amssymb}
\usepackage{soul}
\usepackage[version=3]{mhchem}

\usepackage{color}
\setulcolor{red}

\begin{document}
\title{Multivalley dark solitons in multicomponent Bose-Einstein condensates with repulsive interactions}
\author{Yan-Hong Qin$^{1,3}$}
\author{Li-Chen Zhao$^{1,2,3}$}\email{zhaolichen3@nwu.edu.cn}
\author{Zeng-Qiang Yang$^{4}$}
\author{Liming Ling$^{5}$}\email{linglm@scut.edu.cn}
\address{$^{1}$School of Physics, Northwest University, Xi'an 710127, China}
\address{$^{2}$NSFC-SPTP Peng Huanwu Center for Fundamental Theory, Xi'an 710127, China}
\address{$^{3}$Shaanxi Key Laboratory for Theoretical Physics Frontiers, Xi'an 710127, China}
\address{$^{4}$Department of Physics, School of Arts and Sciences, Shaanxi University of Science and Technology, Xi'an 710021, China}
\address{$^{5}$School of Mathematics, South China University of Technology, Guangzhou 510640, China}
 %%%%%%%%%%%%%%%%%%%%%%%%%%%%%%%%%%%%%%%%%%%%%%%%%
 \begin{abstract}
We obtain multivalley dark soliton solutions with asymmetric or symmetric profiles in multicomponent repulsive Bose-Einstein condensates by developing the Darboux transformation method. We demonstrate that the width-dependent parameters of solitons significantly affect the velocity ranges and phase jump regions of multivalley dark solitons, in sharp contrast to scalar dark solitons. For double-valley dark solitons, we find that the phase jump is in the range $[0,2\pi]$, which is quite different from that of the usual single-valley dark soliton. Based on our results, we argue that the phase jump of an $n$-valley dark soliton could be in the range $[0,n\pi]$, supported by our analysis extending up to five-component condensates. The interaction between a double-valley dark soliton and a single-valley dark soliton is further investigated, and we reveal a striking collision process in which the double-valley dark soliton is transformed into a breather after colliding with the single-valley dark soliton. Our analyses suggest that this breather transition exists widely in the collision processes involving multivalley dark solitons. The possibilities for observing these multivalley dark solitons in related Bose-Einstein condensates experiments are discussed.

\end{abstract}
\date{\today}

\maketitle

\section{Introduction}

Multicomponent Bose-Einstein condensates (BECs) provide a good platform for the investigation of vector solitons both theoretically and experimentally \cite{BEC1,BEC2,BEC3} due to the abundance of intra- and interatomic interactions. Various vector solitons have been investigated in multicomponent BECs with attractive or repulsive interactions \cite{BEC1,BEC2,BS1,BS2,BB,DD1,DD2,DD3,D-antiD,spinzhao,DB1,DB2,DB3,DB4,DDB,DBB}. The major theme of research on attractive BECs is bright solitons \cite{BEC1,BS1,BS2,BB}, while studies on dark solitons (i.e., bright-dark solitons) are considerably hampered by their background modulation instability \cite{MI}. This characteristic makes it difficult to observe dark solitons experimentally in attractive BECs.

In contrast, many more dark vector solitons have been experimentally observed in multicomponent repulsive BECs, such as dark-dark solitons \cite{DD1,DD2,DD3}, dark-bright solitons \cite{DB1,DB2,DB3,DB4}, dark-antidark solitons \cite{D-antiD}, dark-dark-bright solitons and dark-bright-bright solitons \cite{DDB}. Very recently, experimental observations of the collisions of bright-dark-bright solitons were realized in three-component BECs with repulsive interactions \cite{DBB}. Nevertheless, the dark solitons in the abovementioned works refer mainly to single-valley dark solitons (SVDSs). Therefore, we aim to look for multivalley dark soliton (MVDS) solutions in repulsive BECs.

%considering the close relations between soliton solutions in attractive and repulsive systems \cite{MDS2}.

In this work, we present the exact MVDS solutions in multicomponent BECs with repulsive interactions by further developing the Darboux transformation (DT) method. The explicit soliton solutions admit an MVDS in one of the components and multihump bright solitons in the other components. In particular, the soliton width-dependent parameters have a considerable impact on both the velocity range and the phase jump of the MVDS. The phase jump of a double-valley (triple-valley) dark soliton can vary in the range of $[0,2\pi]$ ($[0,3\pi]$). These characteristics of MVDS are distinct from those of the well-known scalar dark soliton, for which the width that depends on both the velocity and the phase jump can be varied in the range $[0,\pi]$\cite{TVPzhao,DS1,DS2}. Furthermore, we explore the collision dynamics of MVDSs. The interaction between two MVDSs reflects the density profile variations only after a collision. Interestingly, one MVDS can transition to a breather after colliding with an SVDS; this can occur because the mixture of the effective energies of solitons in the three components emerges during the collision process. This breather transition occurs extensively in collision processes involving MVDSs.  These findings provide an important supplement for recent reports on nondegenerate vector solitons \cite{BB2,Stalin1,Stalin2,Stalin3}. We expect that more abundant MVDSs could exist in coupled BECs comprising more components and that the phase jump of an $n$-valley dark soliton could be in the range of $[0,n\pi]$.

The remainder of this paper is organized as follows. In Sec.\Rmnum{2}, we introduce the theoretical model and present the double-valley dark soliton (DVDS) solutions in three-component repulsive BECs; we further show the density profiles and analyze the phase features of DVDSs. In Sec.\Rmnum{3}, we investigate the collision dynamics of DVDSs and report the striking state transition dynamics when they collide with SVDSs. In Sec.\Rmnum{4}, we extend our study to four-component repulsive BECs, where triple-valley dark solitons (TVDSs) can be obtained. Finally, the conclusions and discussion are presented in Sec.\Rmnum{5}.

\section{Double-valley dark solitons in three-component repulsive condensates}
\subsection{Physical model and double-valley dark soliton solutions}
We note that it is difficult to obtain dark solitons with more than a single valley in $N (N\leqslant2)$-component nonlinear systems based on the existing  methods, such as the inverse scattering method \cite{DD0,asymptotic1}, DT method \cite{Mat,DT1}, Hirota's method \cite{Hirota1,Hirota2}, Kadomtsev-Petviashvili reduction method \cite{KP1,KP2}, and B\"{a}cklund transformation \cite{BT}. Therefore, we first attempt to find MVDSs in three-component BECs with repulsive interactions. At sufficiently low temperatures and in the framework of the mean-field approach, a three-component repulsive BEC (elongated along the x direction) is well described by the three-component Manakov model \cite{DBB}:
\begin{eqnarray}\label{3nls}
\rm{i}q_{j,t}+\frac{1}{2}q_{j,xx}-(|q_1|^2+|q_2|^2+|q_3|^2)q_j=0,
\end{eqnarray}
where $q_{j}(x,t) (j=1,2,3)$ represents the mean-field wave functions of three-component repulsive BECs. This model can also be used to describe the evolution of light in defocusing nonlinear optical fibers \cite{DS1,DS2}. A recent experiment on bright-dark-bright solitons in repulsive BECs strongly supports the applicability of the above integrable three-component Manakov model to three-component BECs with repulsive interactions \cite{DBB}. For their attractive counterpart, single-hump bright soliton solutions \cite{Hirota1,Lakshman,Mat,Dok} and even multihump bright soliton solutions have been obtained in a multicomponent Manakov system by the DT method \cite{BB2}, Hirota method \cite{Stalin1,Stalin2,Stalin3} and other methods \cite{PCS1,MDS2}. Next, we systematically seek the MVDS solutions in the multicomponent repulsive Manakov model.

The DT method is an effective and convenient way to derive localized wave solutions \cite{lingDS,DT2,DT3,DT4,DT5,DT6}. Recently, it was reported that SVDSs can be obtained through the DT method for multicomponent repulsive Manakov systems \cite{lingDS}. In this paper, we further develop the DT method \cite{lingDS} to derive MVDS solutions in combination with the multifold DT for deriving nondegenerate bright solitons \cite{BB2}. We find that MVDSs can be derived by performing a multifold DT with some special constraint conditions on the eigenfunctions of the Lax pair. For example, one DVDS can be obtained by performing a two-fold DT with the spectral parameters written as $\lambda_j=\frac{1}{2}(\xi_j+\frac{1}{\xi_j})$ and adding some special constraint conditions to the eigenfunctions. The complex parameter $\xi_j=-v_1+\textrm{i}w_j (j=1,2)$ is introduced to simplify the soliton solution and facilitate the analysis of physical meaning of each parameter; the real part determines the soliton's velocity, while the imaginary part is called a soliton width-dependent parameter. The detailed derivation of exact DVDS solutions is given in Appendix A. For the DVDS solutions given in Eq.~\eqref{eq:second transformation}, one DVDS is admitted in the first component, and one double-hump bright soliton is allowed in the other two components (DBBS). By simplifying Eq.~\eqref{eq:second transformation}, the exact general soliton solutions can be expressed as follows:
\begin{subequations} \label{DBB}
\begin{align}
&q_1=\frac{N_1}{M_1}e^{-\textrm{i}t} \label{2a}\\
&q_2=-\textrm{i} 2w_1\sqrt{1\!-\!v_1^2\!-\!w_1^2}\frac{\alpha_1}{\xi_1}\frac{N_2}{M_1}e^{\textrm{i}\left[v_1x-\frac{1}{2}(2+v_1^2-w_1^2)t\right]}\label{2b}\\
&q_3=-\textrm{i} 2w_2\sqrt{1\!-\!v_1^2\!-\!w_2^2}\frac{\beta_2}{\xi_2}\frac{N_3}{M_1}e^{\textrm{i}\left[v_1x-\frac{1}{2}(2+v_1^2-w_2^2)t\right]}\label{2c}
\end{align}
\end{subequations}
with
\begin{equation*}\label{G-dbb}
\begin{aligned}
N_1&\!=\!\left[\frac{\xi_2^*}{\xi_2}\alpha_1^2 e^{\kappa_1-\kappa_2}+\frac{\xi_1^*}{\xi_1}\beta_2^2 e^{\kappa_2-\kappa_1}+\alpha_1^2\beta_2^2e^{\kappa_1+\kappa_2}\right]\\
&\times(w_1+w_2)^2+\frac{\xi_1^*\xi_2^*}{\xi_1\xi_2}(w_1-w_2)^2e^{-\kappa_1-\kappa_2},\\
N_2&\!=\!(w_1+w_2)\left[\beta_2^2(w_1+w_2)e^{\kappa_2}\!+\!(w_1-w_2)e^{-\kappa_2}\right],\\
N_3&\!=\!(w_1+w_2)\left[\beta_2^2(w_1+w_2)e^{\kappa_1}\!+\!(w_2-w_1)e^{-\kappa_1}\right],\\
M_1&\!=\!\left[\alpha_1^2\beta_2^2 e^{\kappa_1+\kappa_2}\!+\!\alpha_1^2e^{\kappa_1-\kappa_2}\!+\!\beta_2^2e^{\kappa_2-\kappa_1}\right]\!\!(w_1+w_2)^2\\
&+(w_1-w_2)^2e^{-\kappa_1-\kappa_2},\\
\kappa_1&=w_1(x-v_1t), \xi_1=-v_1+\textrm{i}w_1, \\
\kappa_2&=w_2(x-v_1t),\xi_2=-v_1+\textrm{i}w_2,\\
\end{aligned}
\end{equation*}
where $\alpha_1,$ $\beta_2$, $v_1$, $w_1$, and $w_2$ are arbitrary real constants. Among them, the parameters $w_1$ and $w_2$ are two soliton width-dependent parameters ($w_1\neq w_2$). The parameter $v_1$ is the soliton velocity, which must satisfy the constraint condition $v_1^2+w_m^2<1$, where $w_m$ is the larger of the two width-dependent parameters, meaning that the soliton width-dependent parameters affect the velocity range. In contrast, the width of a scalar dark soliton depends on the velocity, which cannot exceed the speed of sound \cite{BEC2,TVPzhao,DS1,DS2}. The parameters $\alpha_1$ and $\beta_2$ are two free parameters related to the relative valley values and the center positions of the two valleys; therefore, these parameters nontrivially
contribute to the soliton profiles.

\subsection{Density profiles of double-valley dark solitons}

\begin{figure}[htp]
\centering
\includegraphics[width=85mm]{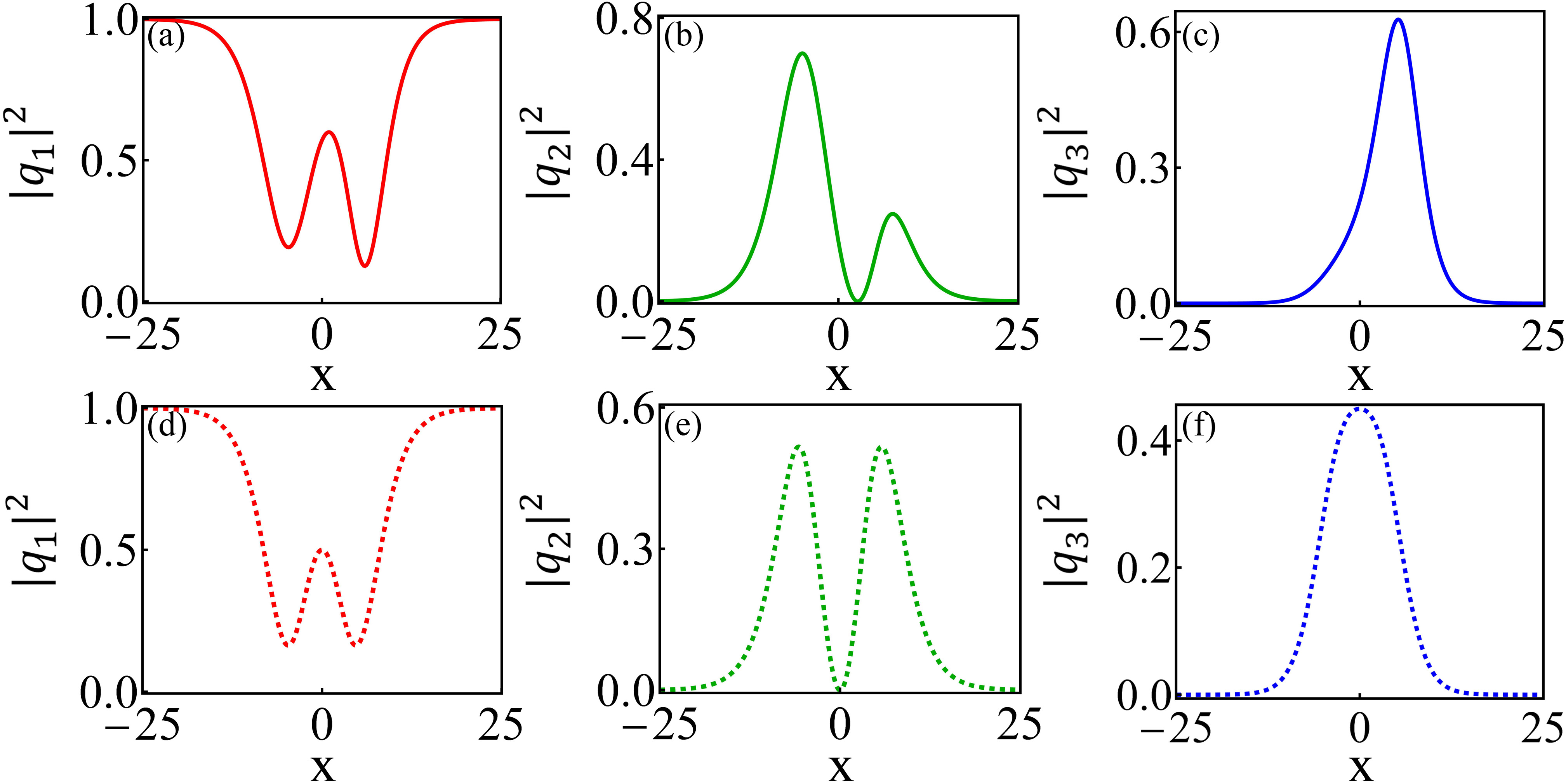}
\caption{The density profiles of DVDSs. (a) shows an asymmetric DVDSs in component $q_1$. (b) and (c) show the density profiles of asymmetric bright solitons in components $q_2$ and $q_3$, respectively. (d-f) show the corresponding profiles for symmetric solitons. The parameters are $ \alpha_1=0.5, \beta_2=0.2$ for asymmetric solitons (a-c) and $ \alpha_1=\beta_2=1/\sqrt{5}$ for symmetric solitons (d-f). The other parameters are chosen as $v_1=0.1$, $w_1=0.2$, and $w_2=0.3$.}\label{Fig1}
\end{figure}

The solution expressed in Eqs.~\eqref{DBB} is generally asymmetric but becomes symmetric when the two free parameters satisfy the condition $\alpha_1=\beta_2=\sqrt{(w_2-w_1)/(w_1+w_2)}$. Examples of asymmetric and symmetric DBBS solutions are shown in Fig.~\ref{Fig1} with solid lines and dashed lines, respectively. The plots from left to right correspond to component $q_1$ (red line), component $q_2$ (green line) and component $q_3$ (blue line). For the asymmetric case depicted in Figs.~\ref{Fig1}(a)-(c), the parameters are $v_1=0.1$, $w_1=0.2$, $v_3=0.3$, $\alpha_1=0.5$, and $\beta_2=0.2$. We can see that a DVDS emerges in component $q_1$, manifesting two spatially localized density ``dips'' on a uniform background. An asymmetric double-hump bright soliton with one node is exhibited in component $q_2$, and an asymmetric single-hump bright soliton without a node presents in component $q_3$ \cite{QM}; these two bright solitons are quite different from the fundamental dark-bright-bright soliton observed in \cite{DDB,DBB}. As shown in Fig.~\ref{Fig1}(a), the two valleys of the asymmetric DVDS are not equal in either the width or the valley values, in sharp contrast to dark solitons reported in previous studies \cite{BEC1,BEC2,BD,DD1,DD2,DD3,D-antiD,spinzhao,DB1,DB2,DB3,DB4,DDB,DBB,lingDS,Hirota1,Hirota2,KP1,KP2,
DS1,DS2}. This characteristic is not observed for the stationary symmetric MVDS solution \cite{PCS2}, which was reported to exist in focusing photorefractive media (similar to attractive multicomponent BECs). The bright solitons that present in components $q_2$ and $q_3$ are similar to those obtained in a two-component attractive (focusing) system \cite{BB2,Stalin1,Stalin2,Stalin3}.

\begin{figure}[htp]
\includegraphics[height=43 mm]{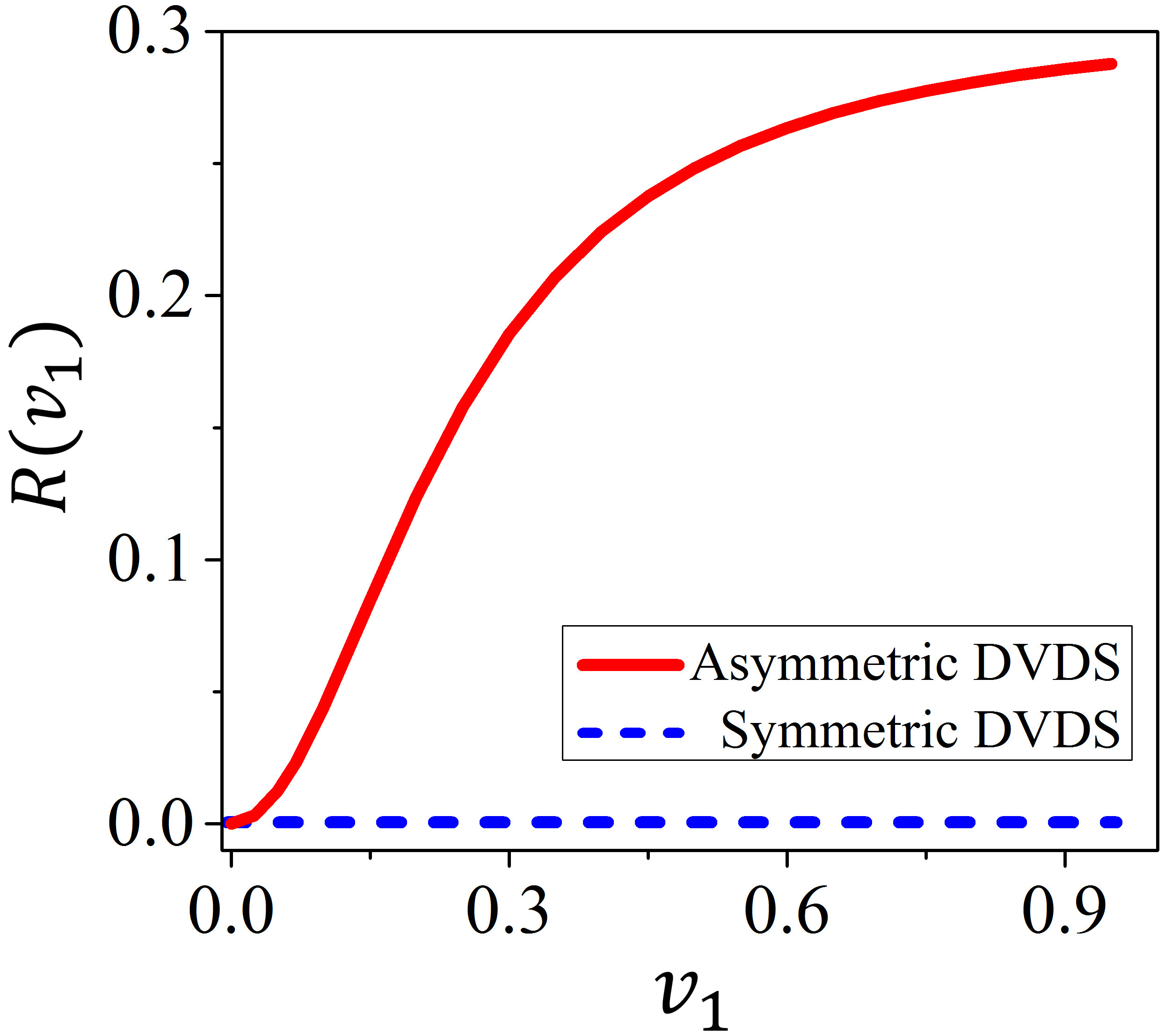}
\caption{The asymmetry degree vs moving velocity for DVDSs. The red solid line and blue dashed line correspond to the asymmetric and symmetric DVDSs, respectively. The asymmetry degree increases with increasing moving velocity for asymmetric DVDSs. However, that for symmetric DVDSs remains symmetric with varying velocity. The other parameters are the same as in Fig. \ref{Fig1}.}\label{Fig2}
\end{figure}
By setting the parameters $\alpha_1=\beta_2=\frac{\sqrt{w_2-w_1}}{\sqrt{w_2+w_1}}$ and the other parameters the same as in the asymmetric case, the solution given in Eqs.~\eqref{DBB} admits symmetric soliton profiles, as shown in Figs.~\ref{Fig1}(d)-(f) (dashed lines). These profiles are similar to those reported in previous studies \cite{MDS2,PCS2}, but the total density can differ from the ``sech-squared'' form. Moreover, the DVDS is expressed by exponential functions, which is distinctly different from the expressions with associated Legendre polynomials \cite{PCS2}. Moreover, the above DVDS solution is more general than those given in \cite{PCS2,MDS2} since it admits more free physical parameters. The soliton velocity of the solution in Eqs.~\eqref{DBB} changes in the range of $\big[0,\sqrt{1-w_m^2}\big)$. In particular, the interaction between the solitons can be investigated analytically with further iterations by applying the developing DT method. When the soliton width-dependent parameters $w_1$ and $w_2$ are close to each other, both components $q_2$ and $q_3$ can show a symmetric (or an asymmetric) double-hump bright soliton.
The characteristics of bright solitons are similar to the results in a two-component attractive (focusing) system \cite{BB2,PCS1,Stalin1,Stalin2,Stalin3}.
In fact, nondegenerate two-component bright soliton solutions can be reduced from the above three-component solution with the aid of the close relations between soliton solutions in attractive and repulsive systems \cite{MDS2}.

It is well known that the valley depth of an SVDS tends to decrease with increasing moving velocity, and the symmetric properties do not change. A similar characteristic is admitted by symmetric DVDSs. However, we find that this characteristic no longer holds for asymmetric DVDSs, whose asymmetry can significantly change with variation in the moving velocity, and the valley depth also tends to become shallower with increasing moving velocity. The asymmetry degree is defined as
\begin{eqnarray}
R(v_1)=\frac{I_1-I_2}{I_1+I_2},
\end{eqnarray}
where $I_1$ and $I_2$ are the values of the left and right valleys in space, respectively. We exhibit the asymmetry degree vs the moving velocity in Fig.~\ref{Fig2} while keeping the other parameters the same as in Fig.~\ref{Fig1}. The red solid curve (blue dashed line) corresponds to the asymmetric (symmetric) DVDS. The asymmetry degree crucially depends on the velocity for asymmetric DVDSs, whereas $R(v_1)$ is constant  for  symmetric DVDSs. This is also a significant difference between symmetric and asymmetric DVDSs.

\subsection{Phase properties of double-valley dark solitons}

\begin{figure}[htp]
  \centering
  \includegraphics[width=82mm]{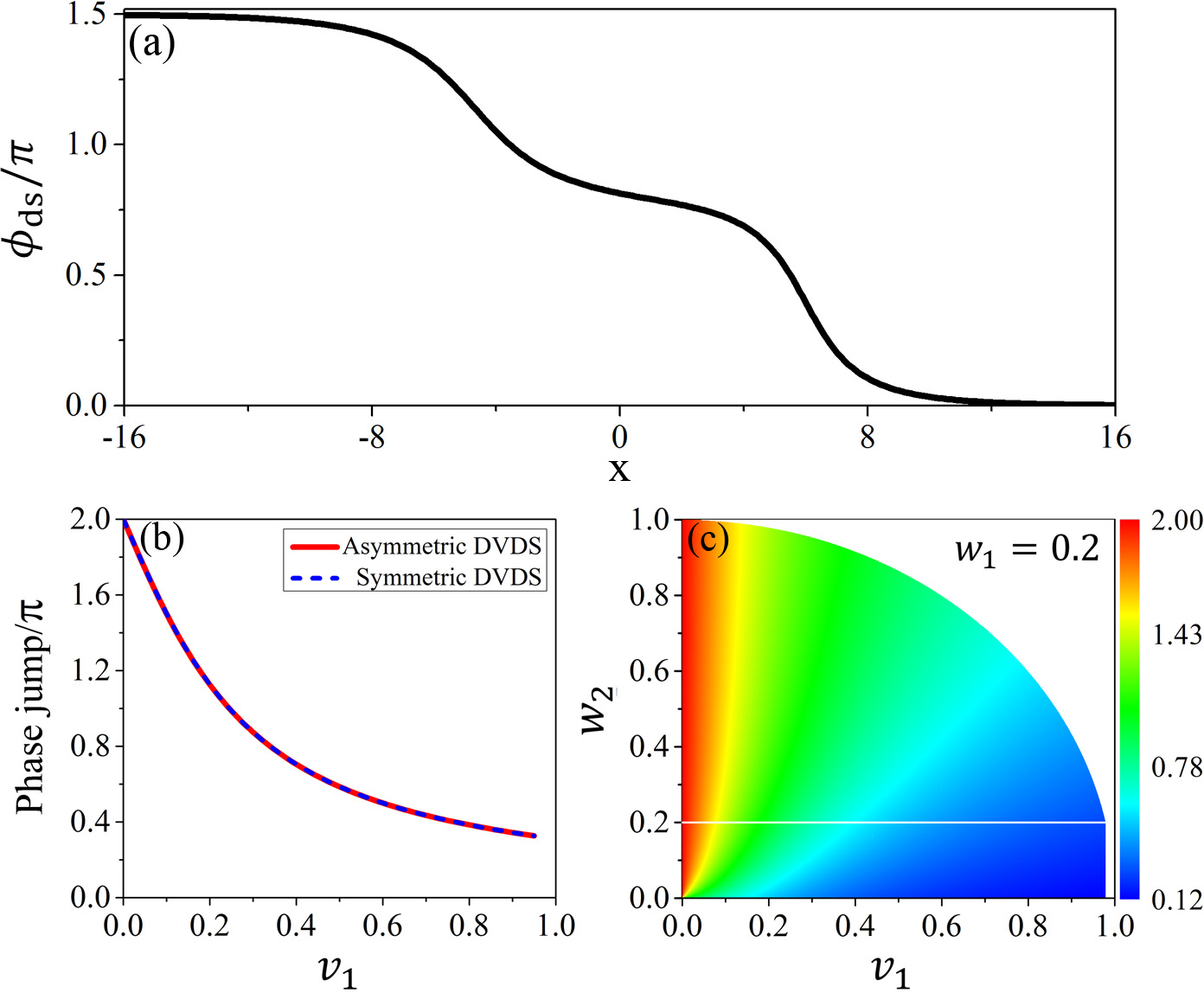}\\
\caption{The phase properties of a DVDS. (a) The phase distribution of an asymmetric DVDS corresponding to the density profile in Fig.~\ref{Fig1}(a). We can see that the DVDS admits two phase jumps across two valleys. (b) The phase jump vs moving velocity for DVDSs. The red solid line shows the curve for the asymmetric DVDS, while the blue dashed line shows the curve for the symmetric DVDS. The other parameters for both cases are identical to those in Fig.~\ref{Fig1}. The phase jump decreases gradually with increasing velocity. (c) The phase jump (in $\pi$ unit) vs the soliton width-dependent parameter and velocity. The parameter $w_1$ is fixed at $0.2$. This demonstrates that the soliton width-dependent parameters significantly affect the phase jump and velocity region of DVDSs. The other parameters are $\alpha_1=0.5$ and $\beta_2=0.2$.}\label{Fig3}
\end{figure}

SVDSs are well known to admit one phase jump across the dip, which varies in the range $[0,\pi]$ \cite{BEC2,BEC1,DS1,DS2,TVPzhao}. Here, we characterize the phase jump properties of DVDSs. Generally, the phase value can be obtained by calculating the argument $\phi_{ds}(x)=\rm{Arg}(q_{ds})$ by dropping the momentum phase of the background. It should be emphasized that the phase distribution in space cannot be directly obtained by calculating only the argument based on the real and imaginary parts of the wave function. Importantly, the phase jump direction must be judged by the phase gradient flow $\partial_x \arctan\left[\frac{\rm{Im }(q_{ds})}{\rm{Re}(q_{ds})}\right]$. Namely, positive (negative) phase gradient flow corresponds to a rising (falling) phase jump. Here, we define the phase jump as $\triangle\phi_{ds}=\phi_{ds}(x\rightarrow-\infty)-\phi_{ds}(x\rightarrow+\infty)$. As an example, we plot the corresponding phase distribution of Fig.~\ref{Fig1}(a) in Fig.~\ref{Fig3}(a). Remarkably, the DVDS is characterized by two phase jumps through two valleys, in contrast to the SVDS with one phase jump \cite{BEC1,BEC2,BD,DD1,DD2,DD3,D-antiD,spinzhao,DB1,DB2,DB3,DB4,DDB,DBB,lingDS,Hirota1,Hirota2,KP1,KP2,
DS1,DS2, TVPzhao}. Moreover, the phase jump value across a DVDS is larger than $\pi$ in this case, which is also distinctive from the phase jump of a SVDS (which cannot exceed $\pi$) \cite{BEC1,BEC2,BD,DD1,DD2,DD3,D-antiD,spinzhao,DB1,DB2,DB3,DB4,DDB,DBB,lingDS,Hirota1,Hirota2,KP1,KP2,
DS1,DS2, TVPzhao}. We note that SVDSs can admit a phase jump greater than $\pi$ in a saturable self-defocusing material \cite{darker1,darker2} but only one phase jump across the soliton. The phase distribution of a symmetric DVDS is similar to that of an asymmetric DVDS, with only a small difference in the spatial position. Additionally, a bright soliton with one node in component $q_2$ always maintains a $\pi$ phase jump, where the abrupt phase change occurs at the node, while a bright soliton without a node has no phase jump in component $q_3$.

Furthermore, we find that the phase jump of a DVDS changes with variation in the moving velocity. As an example, we display the variation in the phase jump of a DVDS with the velocity in Fig.~\ref{Fig3}(b), where the red solid line and the blue dashed line represent asymmetric and symmetric DVDSs, respectively. The parameters are the same as in Fig.~\ref{Fig1} except for the velocity $v_1$. The phase jump value decreases dramatically with increasing velocity. For this case, the velocity range is $|v_1|\in[0,\sqrt{91}/10)$, and the phase jump is within the range $[0.32\pi,2\pi]$. Under the limit of zero velocity, i.e., for a stationary DVDS, the phase jump takes a value of $2\pi$. Fig.~\ref{Fig3}(b) suggests that the change in the phase jump with velocity for a symmetric DVDS is identical to that for an asymmetric DVDS, and the parameters $\alpha_1$ and $\beta_2$ do not affect the phase jump region. However, further analysis indicates that the soliton width-dependent parameters essentially determine the phase jump range. For example, we draw the variation in the phase jump value with the soliton width-dependent parameter $w_2$ in Fig.~\ref{Fig3}(c), where $w_1$ is fixed to be $0.2$ herein. With varying $w_2$, we can calculate the velocity region under the constraint condition $w_j^2+v_1^2<1$. Namely, when $w_2\in(0,0.2)$,  $|v_1|\in[0,2\sqrt{6}/5)$; additionally, when $w_2\in(0.2,1)$,  $|v_1|\in[0,\sqrt{1-w_2^2})$. Fig.~\ref{Fig3}(c) clearly shows the variation in the phase jump with an increase in the width-dependent parameter $w_2$, always taking a value of $2\pi$ for a static DVDS. With an increase in the width-dependent parameter $w_2$, the phase jump region decreases since the phase jump value increases at the maximum velocity. With $w_1= 0.2$, the phase jump range is
$[0.12\pi,2\pi]$. For smaller soliton width-dependent parameters $w_1$ and $w_2$, the phase jump value tends to be zero when the soliton's velocity approaches the limit of the maximum velocity, meaning that DVDSs admit the largest phase jump range of $[0,2\pi]$.

We emphasize that the soliton width-dependent parameters also affect the velocity region; as a result, the maximum speed of the DVDS can be much smaller than the speed of sound for DVDSs when choosing larger width-dependent parameters (see Fig.~\ref{Fig3}(c)). This is a notable characteristic for MVDSs that is absent for usual scalar SVDSs \cite{DS1,DS2}. For usual scalar SVDSs, the maximum velocity is the speed of sound. For a bright-dark soliton in a two-component BEC with attractive interactions, the width and velocity are completely independent of each other \cite{BD}, and there is no limit on the moving velocity. However, for dark-bright solitons in repulsive condensates, a change in the soliton width also causes the velocity range to vary \cite{BD}. Our detailed analysis indicates that the relation between the soliton width and velocity satisfies an inequality for MVDSs and dark-bright solitons \cite{BD}, but this relation becomes an equality for usual scalar SVDSs \cite{DS1,DS2}. This characteristic of dark-bright solitons has not been taken seriously in previous work \cite{DB2}. Nevertheless, we expect that the soliton width plays an important role in the motion of dark-bright solitons in harmonic traps, in contrast to the motion of scalar dark solitons in harmonic traps \cite{Busch}. Therefore, the effects of the soliton width should be considered when discussing the motion of MVDSs in external potentials.

Very recently, three-component vector solitons and their collisions were experimentally observed in BECs with repulsive interactions \cite{DBB}. Motivated by these results and the developed density and phase modulation techniques in ultracold atomic systems \cite{DB1,DDB,DBB,DDE,BBC}, we analytically investigate the collision dynamics of DVDSs by performing further iterations of the developed DT method.

\section{Collision dynamics of double-valley dark solitons}

The collision dynamics of DVDSs mainly include two cases: (i) a collision between a DVDS and an SVDS, (ii) a collision between two DVDSs. We investigate these two types of collisions based on two dark soliton solutions derived by the three-fold and four-fold DT (see details in Appendix A). A collision between two DVDSs usually makes each soliton's profile vary, similar to a collision between nondegenerate bright solitons \cite{BB2}. However, the collision between a DVDS and an SVDS demonstrates a striking state transition process: the DVDS transforms into a breather after colliding with the SVDS, and the latter does not admit any oscillation, just a profile variation.

\begin{figure}[htp]
  \centering
  \includegraphics[width=82mm]{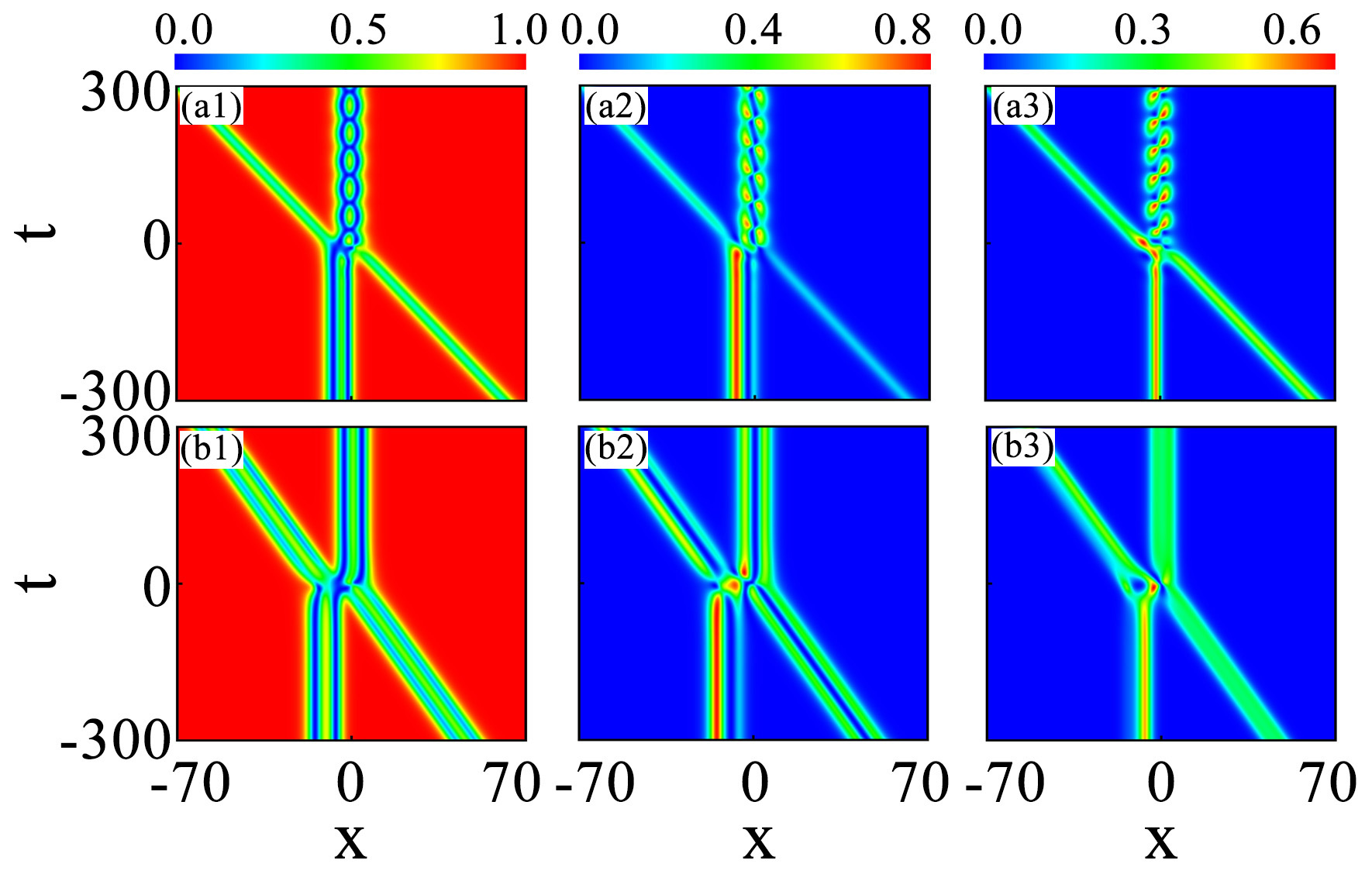}\\
\caption{The collision dynamics of DVDSs. (a1): Collision between one DVDS and an SVDS. There is a striking state transition process in which a DVDS transitions to a breather after the collision, whereas the SVDS does not breath and experiences only slight profile variation. (a2) and (a3) show similar state transition dynamics in the other bright soliton components. (b1): Collision between two DVDSs. For this case, there
is no state transition occurrence for the DVDSs. (b2) and (b3) display similar collision dynamics in the other bright soliton components. The parameters are $v_1=0$, $w_1=0.35$, $w_2=0.6$, $v_2=-0.2$, $w_3=0.3$, $\alpha_1=1$, $\beta_2=1$, and $\alpha_3=\beta_3=1$ for (a1-a3). The parameters are $v_1=-0.15$, $w_1=0.3$, $w_2=0.4$, $v_2=0$, $w_3=0.35$, $w_4=0.45$, $\alpha_1=1$, $\beta_2=1$, $\alpha_3=1$, and $\beta_4=1$ for (b1-b3).}\label{Fig4}
\end{figure}

We first study the interaction between a DVDS and an SVDS based on the exact solution Eq.~\eqref{eq:third transformation} by implementing a three-fold DT with the introduced complex parameters $\xi_1=-v_1+iw_1,\xi_2=-v_1+iw_2$ (generating a DVDS in component $q_1$) and $\xi_3=-v_2+iw_3$ (generating an SVDS in component $q_1$) (see Appendix A for the detailed calculation process). A typical example of the striking state transition process is illustrated in Fig.~\ref{Fig4}(a1). The asymmetric DBBS is evidently transformed into a breather, whose density evolution admits a periodic oscillation behavior. In contrast, the fundamental DBBS (moving toward the left) maintains a soliton state with slight profile variation. Such a state transition induced by a collision is not discussed in the previous literature \cite{DD1,DS1,DS2,DB2,DB1}. Figs.~\ref{Fig4}(a2) and (a3) show the density evolution of bright solitons in components $q_2$ and $q_3$, respectively, demonstrating that this breathing behavior also emerges after the interaction.

To understand the state transition phenomenon, we further analyze the exact solution given by Eq.~\eqref{eq:third transformation} by using the asymptotic analysis technique \cite{asymptotic1,asymptotic2,asymptotic3}. Our analysis suggests that the state transition is induced by the mixture of effective energies of the solitons in the three components during the collision process. Before the collision (within the limit $t\rightarrow-\infty$, with $v_1>v_2$), the DVDS-related vector soliton takes the following asymptotic forms:
\begin{subequations} \label{DBBb}
\begin{align}
&q_{1}^i=\frac{f_1e^{-\nu_1}+f_2e^{\nu_1}+f_3e^{-\nu_2}+f_4e^{\nu_2}}
{m_1e^{-\nu_1}+m_2e^{\nu_1}+m_3e^{-\nu_2}+m_4e^{\nu_2}}e^{-\textrm{i}t},\\
&q_{2}^i=\frac{g_1(g_2e^{-\kappa_2}+g_3e^{\kappa_2})}
{m_1e^{-\nu_1}+m_2e^{\nu_1}+m_3e^{-\nu_2}+m_4e^{\nu_2}}e^{\textrm{i}\varphi_1},\\
&q_{3}^i=\frac{h_1(h_2e^{-\kappa_1}-h_3e^{\kappa_1})}
{m_1e^{-\nu_1}+m_2e^{\nu_1}+m_3e^{-\nu_2}+m_4e^{\nu_2}}e^{\textrm{i}\varphi_2}.
\end{align}
\end{subequations}
After the collision (in the limit $t\rightarrow+\infty$), the asymptotic analysis expressions for the vector soliton take the following forms:
\begin{small}
\begin{subequations} \label{DBBa}
\begin{align}
&q_{1}^f\!=\!\frac{\delta_1\!\left(\delta_2e^{-\textrm{i}\theta}\!+\!\delta_3e^{\textrm{i}\theta}\right)\!+\!\delta_4(\!\delta_5e^{\nu_1}\!+\!\delta_6e^{-\nu_1}\!)\!+\!\delta_7(\!\delta_8e^{\nu_2}\!+\!\delta_9e^{-\nu_2}\!)}
{\varrho_1(\varrho_2e^{-\textrm{i}\theta}\!+\!\varrho_2^*e^{\textrm{i}\theta})\!+\!\varrho_3(\!\varrho_4e^{-\nu_1}\!+\!\varrho_5e^{\nu_1}\!+\!\varrho_6e^{\nu_2}\!+\!\varrho_7e^{-\nu_2}\!)}e^{-\textrm{i}t}, \\
&q_{2}^f\!=\!\frac{e^{\textrm{i}\varphi_1}\zeta_1(\zeta_2e^{-\kappa_2}+\zeta_3e^{\kappa_2})+e^{\textrm{i}\varphi_2}\zeta_4(\zeta_5e^{-\kappa_1}+\zeta_6e^{\kappa_1})}
{\varrho_1(\varrho_2e^{-\textrm{i}\theta}\!+\!\varrho_2^*e^{\textrm{i}\theta})\!+\!\varrho_3(\!\varrho_4e^{-\nu_1}\!+\!\varrho_5e^{\nu_1}\!+\!\varrho_6e^{\nu_2}\!+\!\varrho_7e^{-\nu_2}\!)} , \\
&q_{3}^f=\!\frac{e^{\textrm{i}\varphi_1}\varsigma_1(\varsigma_2e^{-\kappa_2}+\varsigma_3e^{\kappa_2})+e^{\textrm{i}\varphi_2}\varsigma_4(\varsigma_5e^{-\kappa_1}+\varsigma_6e^{\kappa_1})}
{\varrho_1(\varrho_2e^{-\textrm{i}\theta}\!+\!\varrho_2^*e^{\textrm{i}\theta})\!+\!\varrho_3(\!\varrho_4e^{-\nu_1}\!+\!\varrho_5e^{\nu_1}\!+\!\varrho_6e^{\nu_2}\!+\!\varrho_7e^{-\nu_2}\!)}.
\end{align}
\end{subequations}
\end{small}
In  Eqs.~\eqref{DBBb} and Eqs.~\eqref{DBBa},
\begin{eqnarray*}
&&\nu_1=\kappa+\kappa_2,\nu_2=\kappa_1-\kappa_2,\theta=\varphi_1-\varphi_2,\\
&&\kappa_1=w_1(x-v_1t),\varphi_1=v_1x-\frac{1}{2}(v_1^2-w_1^2+2)t,\\
&&\kappa_2=w_2(x-v_1t),\varphi_2=v_1x-\frac{1}{2}(v_1^2-w_2^2+2)t.
\end{eqnarray*}
In addition, the expressions for $f_j$, $g_j$, $h_j$, $m_j$, $\delta_j$, $\zeta_j$, $\varsigma_j$, and $\varrho_j$ are given in Appendix B, and they are all complex constants related to $\xi_j (j=1,2,3)$. From Eqs.~\eqref{DBBb}, we can see that each initial soliton admits a certain velocity. As shown in \cite{TVPzhao}, a moving soliton solution can be transformed to an eigenstate with an eigenenergy if we choose the soliton center as a frame of reference. Therefore, it is reasonable to define an effective energy for the solitons as $E_j^*=\frac{d\phi_j}{dt}$ (where $\phi_j$ is the phase of the wave function). We know that the effective energy of the soliton in components $q_1$, $q_2$ and $q_3$ is $E_1^*=-1$, $E_2^*=-\frac{1}{2}(v_1^2-w_1^2+2)$ and $E_3^*=-\frac{1}{2}(v_1^2-w_2^2+2)$, respectively. For Eqs.~\eqref{DBBa}, we can see that the effective energies of bright soliton components mix and emerge in the DVDS component after the collision process. The breathing behavior obviously originates from the energy mixing term in $e^{\pm i\theta}$. Namely, the effective energy difference of bright solitons determines the oscillation period, and the period is
\begin{eqnarray}\label{df}
T&=& \frac{2\pi} {|E_2^*-E_3^*|}=\frac{4\pi}{|w_1^2-w_2^2|}.
\end{eqnarray}
The oscillation period is identical among the three components. Based on this discussion, we also revisit the collision dynamics of nondegenerate bright solitons \cite{BB2} and find that an effective energy mixture can also emerge. This means that the abovementioned breather behavior is also observable during the collision between a nondegenerate bright soliton and a degenerate bright soliton.

Next, we investigate the interaction between two DVDSs based on the exact solution given in Eq.~\eqref{eq:fourth transformation} by implementing a four-fold DT with the parameters $\xi_1=-v_1+iw_1$, $\xi_2=-v_1+iw_2$ (which generates one DVDS), $\xi_3=-v_2+iw_3$, and $\xi_4=-v_2+iw_4$ (which generates another DVDS) (see Appendix A for the detailed calculation process). For this case, a typical example of the density distribution is displayed in the second panel of Fig.~\ref{Fig4}, for which (b1), (b2) and (b3) correspond to component $q_1$, component $q_2$ and component $q_3$, respectively. As shown in Fig.~\ref{Fig4}(b1), the collision between the two DVDSs causes their profiles to change, accompanied by a phase shift. However, for this case, there is no state transition occurrence for either DVDS, which is dramatically different from the collision dynamics between the DVDS and SVDS described in Fig.~\ref{Fig4}(a1). Moreover, the effective energy mixture no longer emerges in this case. The inelastic collision dynamics for the other two bright soliton components (see Figs.~\ref{Fig4}(b2) and (b3)) are similar to the those for the two nondegenerate bright solitons that collide in \cite{BB2}.

\section{Triple-valley dark solitons in four-component repulsive condensates}

We now extend our discussion to four-component BECs with repulsive interactions, which are described by the following repulsive four-component Manakov model:
\begin{eqnarray}\label{4nls}
\rm{i}q_{j,t}+\frac{1}{2}q_{j,xx}-(|q_1|^2+|q_2|^2+|q_3|^2+|q_4|^2)q_j=0,
\end{eqnarray}
where $q_j(x,t) (j=1,2,3,4)$ denote the four component fields in BECs. By applying the three-fold DT with $\lambda_j=1/2(\xi_j+1/\xi_j)$ and $\xi_j=-v_1+iw_j$ ($j=1,2,3$), which is similar to the DT presented in Appendix A, we can obtain a class of dark-bright-bright-bright soliton solutions, which admits a TVDS in component $q_1$ and triple-hump bright solitons in the other three components. The exact soliton solutions can be expressed in the following form  (we do not provide the explicit solution process here for brevity):
\begin{small}
\begin{subequations}\label{DBBB}
\begin{align}
&q_1=\frac{1}{\xi_1\xi_2\xi_3}\frac{N_1}{M_1}e^{-\textrm{i}t}, \label{3a}\\
&q_2=-\textrm{i}\frac{2w_1}{\xi_1}\alpha\sqrt{1-v_1^2-w_1^2}\frac{N_2}{M_1}e^{\textrm{i}\left[v_1x-\frac{1}{2}(2+v_1^2-w_1^2)t\right]},\label{3b}\\
&q_3=-\textrm{i}\frac{2w_2}{\xi_2}\beta\sqrt{1-v_1^2-w_2^2}\frac{N_3}{M_1}e^{\textrm{i}\left[v_1x-\frac{1}{2}(2+v_1^2-w_2^2)t\right]},\label{3c}\\
&q_4=-\textrm{i}\frac{2w_3}{\xi_3}\gamma\sqrt{1-v_1^2-w_3^2}\frac{N_4}{M_1}e^{\textrm{i}\left[v_1x-\frac{1}{2}(2+v_1^2-w_3^2)t\right]},\label{3c}
\end{align}
\end{subequations}
\end{small}
with
\begin{small}
\begin{equation*}\label{G-dbb}
\begin{aligned}
\kappa_1=&2w_1(x-v_1t),\kappa_2=2w_2(x-v_1t),\kappa_3=2w_3(x-v_1t),\\
N_1\!=&\xi_1\xi_2^*\xi_3^*\alpha^2\eta_2e^{\kappa_1}+\xi_1^*\xi_2\xi_3^*\beta^2\eta_3e^{\kappa_2}+\xi_1^*\xi_2^*\xi_3\gamma^2\eta_4e^{\kappa_3}+\\
&\eta_5\!\Big[\xi_1\xi_2\xi_3\alpha^2\beta^2\gamma^2e^{\kappa_1+\kappa_2+\kappa_3}\!+\!\xi_1^*\xi_2\xi_3\beta^2\gamma^2e^{\kappa_2+\kappa_3}\!+\!\\
&\xi_1\xi_2\xi_3^*\alpha^3\beta^2e^{\kappa_1+\kappa_2}\!+\!\xi_1\xi_2^*\xi_3\alpha^2\gamma^2e^{\kappa_1+\kappa_3}\!\Big]\!+\!\xi_1^*\xi_2^*\xi_3^*\eta_1,\\
N_2\!=&\Big[\rho_1+\rho_2\beta^2e^{\kappa_2}+\rho_3\gamma^2e^{\kappa_3}+\eta_5\beta^2\gamma^2e^{\kappa_2+\kappa_3}\Big]e^{\kappa_1/2},\\
N_3\!=&\Big[\rho_4+\rho_5\alpha^2e^{\kappa_1}-\rho_3\gamma^2e^{\kappa_3}+\eta_5\alpha^2\gamma^2e^{\kappa_1+\kappa_3}\Big]e^{\kappa_2/2},\\
N_4\!=&\Big[\rho_6-\rho_5\alpha^2e^{\kappa_1}-\rho_2\beta^2e^{\kappa_3}+\eta_5\alpha^2\beta^2e^{\kappa_1+\kappa_2}\Big]e^{\kappa_3/2},\\
M_1\!=&\Big[\alpha^2\beta^2e^{\kappa_1+\kappa_2}\!+\!\alpha^2\gamma^2e^{\kappa_1+\kappa_3}\!+\!\beta^2\gamma^2e^{\kappa_2+\kappa_3}\Big]\eta_5\!+\!\eta_1\\
&\!+\!\eta_2\alpha^2e^{\kappa_1}\!+\!\eta_3\beta^2e^{\kappa_2}\!+\!\eta_4\gamma^2e^{\kappa_3}\!+\!\eta_5\alpha^2\beta^2\gamma^2e^{\kappa_1\!+\!\kappa_2\!+\!\kappa_3}.\\
\end{aligned}
\end{equation*}
\end{small}
In the expressions listed above, $\eta_1=(w_1\!-\!w_2)^2(w_1\!-\!w_3)^2(w_2\!-\!w_3)^2$, $\eta_2=(w_1\!+\!w_2)^2(w_2\!-\!w_3)^2(w_1\!+\!w_3)^2$,
$\eta_3=(w_1+w_2)^2(w_1-w_3)^2(w_2+w_3)^2$, $\eta_4=(w_1-w_2)^2(w_2+w_3)^2(w_2+w_3)^2$,
$\eta_5=(w_1+w_2)^2(w_1+w_3)^2(w_2+w_3)^2$, $\rho_1=(w_1^2-w_2^2)(w_1^2-w_3^2)(w_2-w_3)^2$,
$\rho_2=(w_1+w_2)^2(w_1^2-w_3^2)(w_2+w_3)^2$, $\rho_3=(w_1^2-w_2^2)(w_1+w_3)^2(w_2+w_3)^2$,
$\rho_4=(w_2^2-w_1^2)(w_1-w_3)^2(w_2^2-w_3^2)$, $\rho_5=(w_1+w_2)^2(w_1+w_3)^2(w_2^2-w_3^2)$, and
$\rho_6=(w_1-w_2)^2(w_1^2-w_3^2)(w_2^2-w_3^2)$. The parameters $v_1$, $w_1$, $w_2$, $w_3$, $\alpha$, $\beta$, and $\gamma$ are real constants that co-determine the profile and position of the TVDS. The parameters $w_1$, $w_2$, and $w_3$ are three soliton width-dependent parameters ($w_1\neq w_2\neq w_3$). The parameter  $v_1$ is the soliton velocity, which should satisfy the constraint $v_1^2+w_j^2<1$ $(j=1,2,3)$.  The parameters $\alpha$, $\beta$ and $\gamma$ are three free parameters associated with  the center positions and relative values of the three valleys.  The profile of the TVDS is generally asymmetric but becomes symmetric for certain values of the parameters $\alpha$, $\beta$, and $\gamma$, which is similar to the above three-component case.
\begin{figure}[htp]
\centering
\includegraphics[width=82mm]{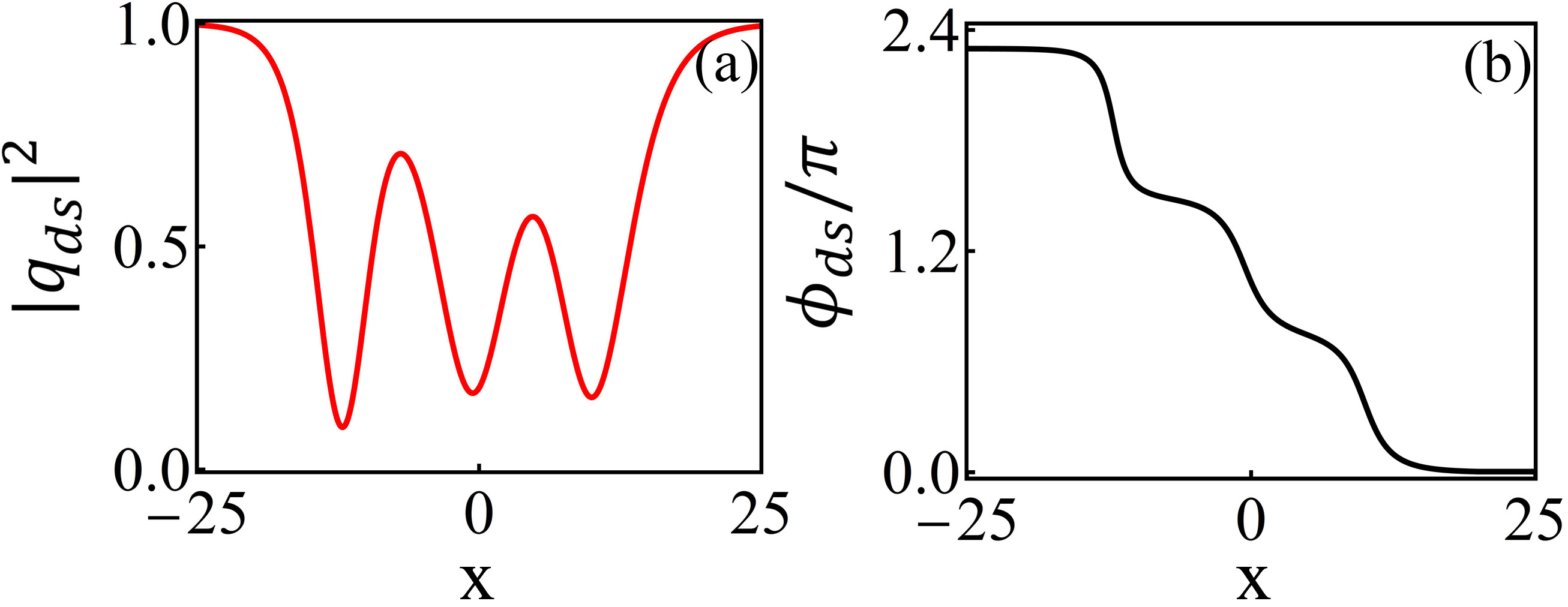}
\caption{The density profile (a) and phase distribution (b) of an asymmetric TVSD. Three phase jumps across the three density valleys can be observed. The parameters are $v_1=0.1$, $w_1=0.2$, $w_2=0.28$, $w_3=0.33$, $\alpha=\frac{1}{6}$, $\beta=\frac{1}{10}$, $\gamma=1$. }\label{Fig5}
\end{figure}
 With the arbitrary setting of these parameters, the solution given in Eqs.~\eqref{DBBB} shows a TVDS  in component $q_1$, a triple-hump bright soliton with two nodes in component $q_2$, a bright soliton with one node in component $q_3$ and a bright soliton without nodes in component $q_4$. The bright solitons in the last three components are similar to the solitons reported in three-component attractive BECs \cite{BB2}. In this section, we mainly discuss the TVDS in the first component.

\begin{figure}[htp]
  \centering
  \includegraphics[width=50mm]{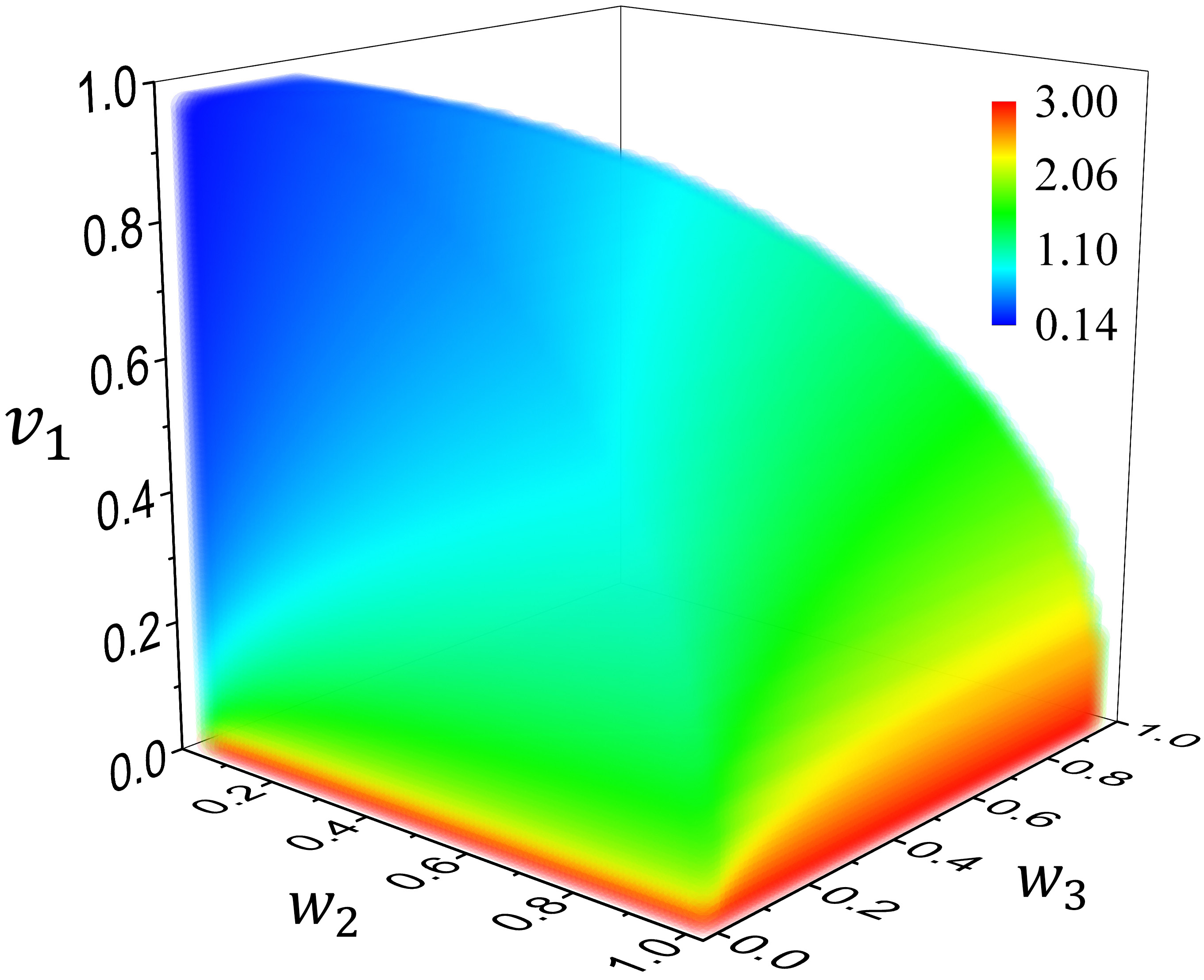}
\caption{The phase jump (in $\pi$ unit) of a TVDS with varying velocity and soliton width-dependent parameters $w_2$ and $w_3$. The width-dependent parameters significantly affect the phase jump and velocity region. $w_1$ is fixed to be $0.2$. The other parameters are $\alpha=1/6$, $\beta=1/10$, $\gamma=1$, and $t=0.$}\label{Fig6}
\end{figure}

The density profile and phase distribution of an asymmetric TVDS are displayed in Fig.~\ref{Fig5}. Fig.~\ref{Fig5}(a) depicts the density distribution of an asymmetric TVDS. The density profile of the TVDS is somewhat similar to those of stationary multisoliton complexes on a background \cite{MCDS1,MCDS2}: both are formed as the special nonlinear superpositions of pairs of bright and dark solitons. However, the derivation method for the TVDS is distinctive. Consequently, our developed DT method can be used to derive more general MVDS solutions with moving velocities and investigate their collisions analytically through further iterations. Fig.~\ref{Fig5}(b) demonstrates that the TVDS features three phase jumps across three valleys. The soliton width-dependent parameter $w_j$ also affects the phase jump region of the TVDS. As an example, we demonstrate the variation in the phase jump value with the changes in the soliton width-dependent parameters ($w_2$ and $w_3$) and the velocity in Fig.~\ref{Fig6}. The parameter $w_1$ is fixed to be $0.2$. For the static TVDS solution, the phase jump is $3\pi$. The phase jump value achieves the minimum value when the velocity tends toward the limit of the maximum velocity. With $w_1=0.2$, the phase jump range of the TVDS is $[0.14\pi,3\pi]$, but the phase jump range of the TVDS can vary within $[0,3\pi]$ by further decreasing the value of $w_1$. The phase jumps of triple-hump bright solitons in component $q_2$, component $q_3$, and component $q_4$ are always $2\pi$, $\pi$ and $0$, respectively. Moreover, the parameters $w_j$ can also vary the soliton velocity region for a TVDS. The velocity range is $[0,\sqrt{1-w_{m}^2})$, where $w_{m}$ is the largest of the three width-dependent parameters. These characteristics are similar to those of the DVDS case, which could hold for dark solitons with more valleys.

Moreover, the collision dynamics of TVDSs include three main cases: a collision between a TVDS and an SVDS, a collision between a TVDS and a DVDS, and a collision between two TVDSs. A breather transition also emerges for the TVDS in the first and second cases, whereas in the third case, while the soliton profiles similarly vary after the collision, there is no breather behavior. The oscillation behavior is much more abundant in these cases than in the cases for the above DVDS. We will systemically discuss their properties in the future with the further development of analysis techniques.

\section{Conclusion and discussion}

We obtain exact DVDS and TVDS solutions by further developing the DT method. Their velocity and phase jump characteristics are characterized in detail, and in particular, we demonstrate that changes in the soliton width-dependent parameters have considerable influences on the velocity and phase jump ranges. This finding indicates that the effects of the soliton width should be considered when studying the motion of MVDSs in external potentials. The collision dynamics of the DVDS and TVDS are also discussed. The collisions involving DVDSs are discussed in detail, and we report a striking state transition process in which a DVDS transitions into a breather after colliding with an SVDS due to the mixture of the effective energies of the soliton states in the three components. Furthermore, our analyses suggest that breather transitions exist widely in the collision processes involving MVDSs.

Our discussion can be extended to $N$-component coupled systems. The $(N-1)$-valley dark soliton solution can be obtained by applying the $(N-1)$-fold DT with similar constraint conditions on the eigenfunctions of the Lax pair. The phase jump of the $(N-1)$-valley dark soliton could vary in the region $[0,(N-1)\pi]$. This argument is supported by our calculation up to a five-component case. Nevertheless, further study is needed to learn how to express the analytical solution for arbitrary $N$-component coupled systems. The collision properties between MVDSs and degenerate vector solitons are expected to be much more abundant than those of previously reported vector soliton collisions \cite{Lakshman,DBB,DB2,BD,DT6}. In particular, the state transition between the breather and MVDS could have some important hints for soliton state manipulation fields.

Recently, three-component vector solitons and their collisions were experimentally observed in BECs with repulsive interactions \cite{DBB}. This finding indicates that soliton dynamics could be quantitatively described by the above integrable repulsive three-component Manakov model. Therefore, we discuss the possibilities to observe DVDSs in three-component repulsive BECs in combination with well-developed quantum engineering techniques \cite{DB1,DDB,DBB,DDE,BBC}. Let us consider quasi-one-dimensional elongated BECs of $^{87}\rm{Rb}$. The different components are the magnetic sublevels $m_F=0,\pm1$ of the $F=1$ hyperfine manifold. In the beginning, all atoms are prepared in the $m_F=0$ state. One can use spatial local control beams to transfer atoms from the initial state $m_F=0$ to $m_F=\pm1$. With knowledge of the density and phase given by the solution of Eqs. \eqref{DBB}, one could transfer the atom and imprint phase on them simultaneously to approach the initial state for the DVDS and bright soliton states in the corresponding components. Our numerical simulations suggest that these soliton states are robust against small deviations and noise, suggesting it is possible to observe these solitons experimentally.

\section*{ACKNOWLEDGMENTS}

This work is supported by the National Natural Science Foundation of China (Contract No. 12022513,11775176,12047502),  the Major Basic Research Program of Natural Science of Shaanxi Province (Grant No. 2018KJXX-094), Scientific research program of Education Department of Shaanxi Provience (18JK0098) and Scientific Research Foundation of SUST (2017BJ-30).

\begin{appendix}
\section{Derivation of the double-valley dark soliton solution of equation (1)}

The $N$-component repulsive BEC can be described by the following $N$-coupled Manakov model \cite{lingDS,DBB}:
\begin{equation}\label{n-nls}
  \textrm{i}\mathbf{q}_t+\frac{1}{2}\mathbf{q}_{xx}-\mathbf{q}^{\dag}\mathbf{q}\mathbf{q}=0,
\end{equation}
where
\begin{eqnarray*}
 \mathbf{q}=(q_1,q_2,...q_N)^{\mathsf{T}},
\end{eqnarray*}
which admits the following Lax pair:
\begin{subequations}\label{Laxpair}
\begin{align}
     &\psi_x=U(\lambda;Q)\psi,\label{a}\\
    & \psi_t=V(\lambda;Q)\psi,\label{b}
\end{align}
\end{subequations}
with
\begin{subequations}
\begin{align}
&U=(\textrm{i}\lambda\sigma_3+\textrm{i}Q),\label{a}\\
&V=\left[\textrm{i}\lambda^2\sigma_3+\textrm{i}\lambda  Q-\frac{1}{2}(\textrm{i}\sigma_3Q^2-\sigma_3Q_x)\right]\label{b},
\end{align}
\end{subequations}
where
\begin{eqnarray*}
&&Q=\left[
    \begin{array}{cc}
      0 & -\mathbf{q}^{\dag}\\
      \mathbf{q} & \mathbf{0}_{N\times N} \\
    \end{array}
  \right],~~\sigma_3=\left[
                       \begin{array}{cc}
                         1 & \mathbf{0}_{1\times N} \\
                         \mathbf{0}_{N\times 1} & -\mathbb{I}_{N\times N} \\
                       \end{array}
                     \right],
\end{eqnarray*}
As mentioned in the main text, multivalley dark solitons can be obtained in an $N(N>2)$ component system. As an example, we will take the DVDS solution derivation process in the three-component case to introduce the calculation method for MVDSs.

To obtain the DVDS solutions, we use the following seed solutions:
\begin{eqnarray}
q_{01}=e^{-\textrm{i}t},q_{02}=0,q_{01}=0.
\end{eqnarray}
First, we need to solve the Lax pair equation \eqref{Laxpair} with the above seed solutions. We use the following gauge transformation:
\begin{eqnarray*}
S=\rm{diag}(1,e^{it},1,1),
\end{eqnarray*}
Which converts the variable coefficient differential equation into a constant-coefficient equation. Then, we can obtain
\begin{subequations}
\begin{align}
& \psi_{0,x}=U_0\psi_0, \\
&\psi_{0,t}=\left(\frac{\rm{i}}{2}U_0^2+\lambda U_0+\frac{\rm{i}}{2}\lambda^2\right)\psi_0,
\end{align}
\end{subequations}
where
\begin{eqnarray*}
U_0=\rm{i}
  \begin{bmatrix}
  \lambda&-1&0&0\\
  1&-\lambda&0&0\\
  0&0&-\lambda&0\\
  0&0&0&-\lambda\\
\end{bmatrix},
\end{eqnarray*}
In the following, we consider the property of $U_0$. We can obtain the characteristic equation of matrix $U_0$ at $\lambda=\lambda_j=a_j+\textrm{i}b_j$:
\begin{equation}\label{EV}
    \begin{split}
     \textrm{det}(\textrm{i}\tau_j-U_0)=(\lambda_j+\tau_j)^2(\tau_j^2-\lambda_j^2+1)=0.
    \end{split}
\end{equation}
The eigenvalues of \eqref{EV} are
\begin{eqnarray}\label{EVV}
\tau_{j1}=-\sqrt{\lambda_j^2-1},\tau_{j2}=\tau_{j3}=-\lambda_j,\tau_{j4}=\sqrt{\lambda_j^2-1}.
\end{eqnarray}
To obtain the vector solution of \eqref{Laxpair}, we further diagonalize the matrix $U_0$. Then, we obtain
\begin{subequations}\label{diagonalization}
\begin{align}
     &\phi_x=\widetilde{U}_0\phi, ~~\phi=H^{-1}S\psi,\\
     &\phi_t=\left(\frac{\textrm{i}}{2}\widetilde{U}_0^2+\lambda_j \widetilde{U}_0+\frac{\textrm{i}}{2}\lambda_j^2\right)\phi,
\end{align}
\end{subequations}
where the transformation matrix $H$ can be expressed as the following form:
\begin{small}
\begin{equation*}
H=
\left(\!
  \begin{array}{cccc}
    \lambda_j+1+\tau_{j1} & 0 & 0&  \lambda_j+1+\tau_{j4} \\
     \lambda_j+1-\tau_{j1} & 0 & 0 &  \lambda_j+1-\tau_{j4} \\
    0& 1 & 0 & 0 \\
    0 & 0 & 1 & 0 \\
  \end{array}
\!\right).
\end{equation*}
\end{small}
Thus, we have the vector solution for  \eqref{diagonalization}:
\begin{small}
\begin{equation}\label{eq:eigenfunctions}
  \phi_j\!=\!\left(\!
     \begin{array}{c}
       \phi_{j1} \\
       \phi_{j2}  \\
       \phi_{j3} \\
       \phi_{j4} \\
     \end{array}
  \! \right)\!=\!\left(\!
\begin{array}{r}
c_{j1}\exp{\textrm{i}[\tau_{j1}x+\frac{1}{2}(\lambda_j^2\!+\!2\lambda\tau_{j1}-\tau_{j1}^2)t]}\\
c_{j2}\exp{\textrm{i}[\tau_{j2}x+\frac{1}{2}(\lambda_j^2\!+\!2\lambda\tau_{j2}-\tau_{j2}^2)t]}\\
c_{j3}\exp{\textrm{i}[\tau_{j3}x+\frac{1}{2}(\lambda_j^2\!+\!2\lambda\tau_{j3}-\tau_{j3}^2)t]}\\
c_{j4}\exp{\textrm{i}[\tau_{j4}x+\frac{1}{2}(\lambda_j^2\!+\!2\lambda\tau_{j4}-\tau_{j4}^2)t]}\\
\end{array}
\!\right).
\end{equation}
\end{small}
The coefficients $c_{j1}$, $c_{j2}$, $c_{j3}$, and $c_{j4}$ are arbitrary complex parameters. Then, the special solution of Lax pair \eqref{Laxpair} at $\lambda_j$ can be obtained based on $\psi_j=S^{-1}H\phi_j$:
\begin{small}
\begin{equation}\label{eq:eigenfunctions1}
\psi_j\!=\!\left(\!
         \begin{array}{c}
           \psi_{j1} \\
           \psi_{j2} \\
           \psi_{j3} \\
           \psi_{j4} \\
         \end{array}
       \!\right)\!
       =
       \!
\left(\!
  \begin{array}{c}
    (\!1\!+\!\lambda\!+\!\tau_{j1}) \phi_{j1}\!+\!(\!1\!+\!\lambda\!+\!\tau_{j4}) \phi_{j4}\\
   \left[(\!1\!+\!\lambda\!-\!\tau_{j1}) \phi_{j1}\!+\!(\!1\!+\!\lambda\!-\!\tau_{j4})\phi_{j4}\right]\!e^{-it} \\
    \phi_{j2} \\
    \phi_{j3} \\
  \end{array}
\!\right)\!.\\
\end{equation}
\end{small}
We can see from \eqref{EVV} that the algebraic equation \eqref{EV} has a pair of opposite complex roots, i.e., $\tau_{j1}=-\tau_{j4}$. To obtain the DVDS, we pick only one of them in special solution \eqref{eq:eigenfunctions1}. Namely, we need to let $\phi_{j1}=0$ (i.e., the coefficient $c_{j1}=0$) or  $\phi_{j4}=0$ (i.e., the coefficient $c_{j4}=0$). In this paper, we choose $\phi_{j4}=0$. Then, the above special solutions $\psi_j$ for spectral problem \eqref{Laxpair} at $\lambda_j$ are re-expressed as follows:
\begin{small}
\begin{equation}\label{eq:eigenfunctions2}
\psi_j=\left(
         \begin{array}{c}
           \psi_{j1} \\
           \psi_{j2} \\
           \psi_{j3} \\
           \psi_{j4} \\
         \end{array}
       \right)\\
       =
\left(\!
  \begin{array}{c}
    (1+\lambda+\tau_{j1}) \phi_{j1}\\
   \left[(1+\lambda-\tau_{j1}) \phi_{j1}\right]e^{-it} \\
    \phi_{j2} \\
    \phi_{j3} \\
  \end{array}
\!\right).
\end{equation}
\end{small}
Next, we need to perform a two-fold DT using the special solutions \eqref{eq:eigenfunctions2} to derive the DVDS solutions. First, we perform the first-step iteration by applying the DT in \cite{lingDS} with $\lambda_1=a_1+\textrm{i}b_1$ and constrain the eigenfunction $\psi_{13}=\phi_{12}=0$, i.e., the coefficient $c_{12}=0$ (or the eigenfunction $\psi_{14}=\phi_{13}=0$, i.e., the coefficient $c_{13}=0$):
\begin{eqnarray}\label{eq:first transformation}
\psi[1]&=&T[1]\psi,\,\,\,T[1]=\mathbb{I}+\frac{\lambda_1^*-\lambda_1}{\lambda-\lambda_1^*}
    \frac{\psi_1\psi_1^\dag\Lambda}{\psi_1^\dag\Lambda\psi_1},\nonumber\\
Q[1]&=&Q+(\lambda_1-\lambda_1^*)\left[\sigma_3,\frac{\psi_1\psi_1^\dag\Lambda}
     {\psi_1^\dag\Lambda\psi_1}\right],
\end{eqnarray}
where $\Lambda=\rm{diag}(1,-1,-1,-1)$ and a dagger denotes the matrix transpose and complex conjugate. For the second-step iteration, we employ $\psi_2$, which is mapped to $\psi_2[1]=T[1]|_{\lambda=\lambda_2}\psi_2$ with $\lambda_2=a_2+\textrm{i}b_2$, and we constrain the eigenfunction $\psi_{24}=\phi_{23}=0$, i.e., the coefficient $c_{23}=0$ (or the eigenfunction $\psi_{23}=\phi_{22}=0$, i.e., the coefficient $c_{22}=0$):
\begin{eqnarray}\label{eq:second transformation}
&&\psi[2]=T[2]\psi[1],T[2]=\mathbb{I}\!+\!\frac{\lambda_2^*\!-\!\lambda_2}{\lambda\!-\!\lambda_2^*}
    \frac{\psi_2[1]\psi_2[1]^\dag\Lambda}{\psi_2[1]^\dag\Lambda\psi_2[1]},\nonumber \\
&&Q[2]=Q[1]+(\lambda_2-\lambda_2^*)\left[\sigma_3,\frac{\psi_2[1]\psi_2[1]^\dag\Lambda}
     {\psi_2[1]^\dag\Lambda\psi_2[1]}\right].
\end{eqnarray}
Furthermore, we require that the solitons' velocities in these two iteration processes be equal. Only if the above iterative processes and velocity requirements are met can the first component of solution $Q[2]$ be the DVDS solution. Then, we need to analyze the relationship between the soliton velocity and spectral parameter $\lambda_j$.

We find that the solitons' velocities can be obtained by calculating the following:
\begin{eqnarray}\label{velocity1}
\widetilde{v}_j=\frac{\left[\textrm{Im}(\tau_{j1})-b_j\right]\textrm{Re}(\tau_{j1})-\left[\textrm{Im}(\tau_{j1})+3b_j\right]a_j}{\textrm{Im}(\tau_{j1})+b_j},
\end{eqnarray}
and the inverse soliton width is
\begin{eqnarray}\label{width1}
\widetilde{w}_j=-\left[\textrm{Im}(\tau_{j1})+b_j\right].
\end{eqnarray}
Moreover, the spectral parameters  should satisfy the constraint condition
\begin{eqnarray}\label{constrain1}
 \left[b_j-\textrm{Im}(\tau_{j1})\right]^2+\left[a_j-\textrm{Re}(\tau_{j1})\right]^2-1>0.
\end{eqnarray}
With the spectral parameter $\lambda_j=a_j+\textrm{i}b_j$, the velocity or width of the soliton cannot be directly determined by the real or imaginary part of the spectral parameter. This greatly increases the difficulty of taking the equal velocity in the two-fold DT \eqref{eq:first transformation}-\eqref{eq:second transformation} with different spectral parameters. Accordingly, we express the real and imaginary parts of the spectrum parameter by the velocity $\widetilde{v}_j$ and width $\widetilde{w}_j$. By combining \eqref{velocity1}-\eqref{constrain1} and performing further calculations, we obtain
\begin{eqnarray}\label{spectrum}
a_j=-\frac{\widetilde{v}_j(1+\widetilde{v}_j^2+\widetilde{w}_j^2)}{2(\widetilde{v}_j^2+\widetilde{w}_j^2)},b_j=\frac{\widetilde{w}_j(1-\widetilde{v}_j^2-\widetilde{w}_j^2)}{2(\widetilde{v}_j^2+\widetilde{w}_j^2)},
\end{eqnarray}
and the constrain condition is represented as
\begin{eqnarray}\label{constrain2}
\widetilde{v}_j^2+\widetilde{w}_j^2<1.
\end{eqnarray}
\setcounter{figure}{0}
\renewcommand\thefigure{A\arabic{figure}}
\begin{figure}[htp]
  \centering
  \includegraphics[width=80mm]{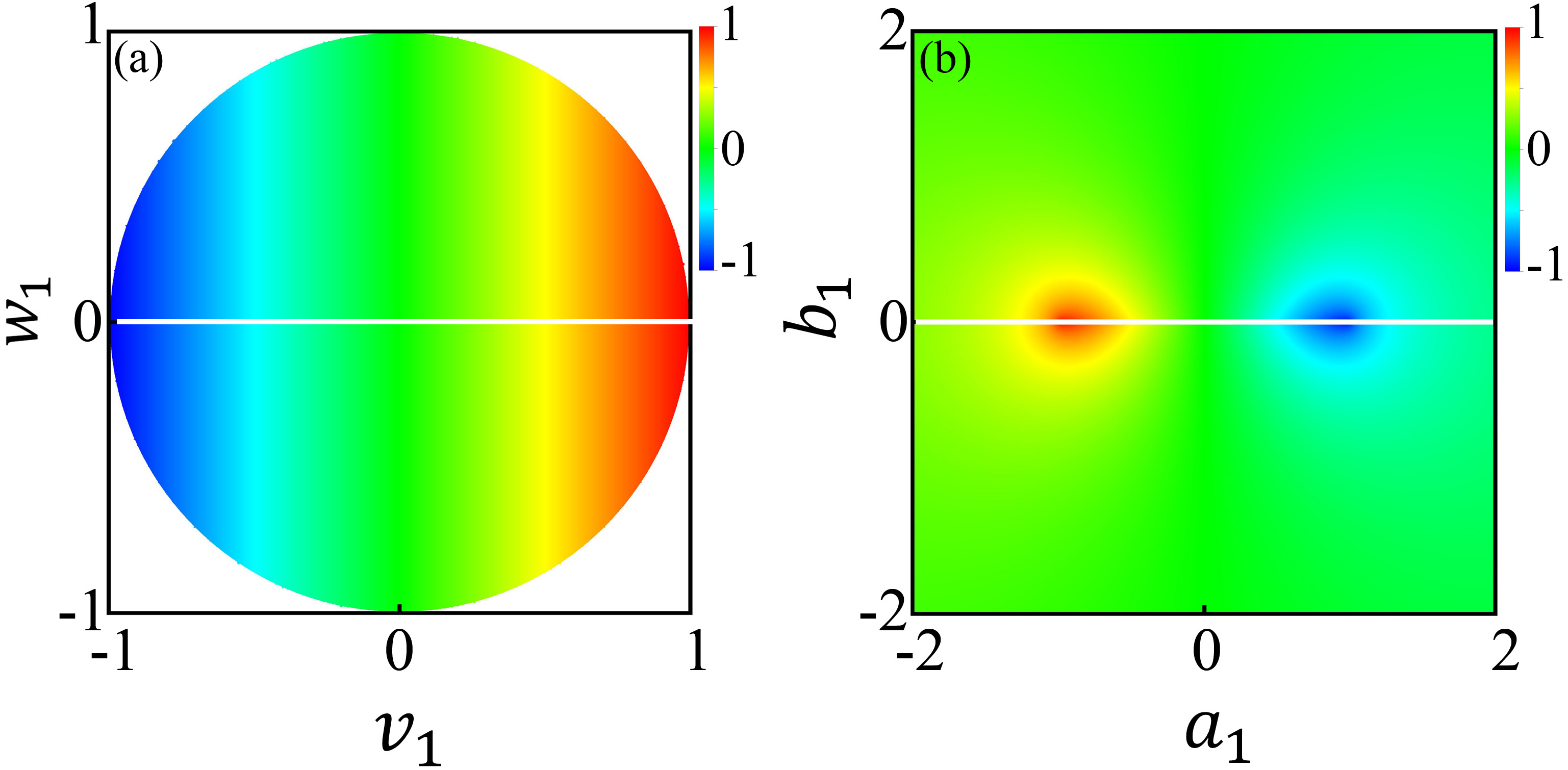}
\caption{The relation between the velocity and spectral parameter expressed by different representations. (a): The spectral parameters are written as $\lambda_1=\frac{1}{2}(\xi_1+\frac{1}{\xi_1})$ and $\xi_1=-v_1+\textrm{i}w_1$. For this case, there is a linear relationship between the velocity and the real part of the spectral parameter; namely, $v_1$ is just simply the velocity of the soliton. (b): The spectral parameter is written as $\lambda_1=a_1+\textrm{i}b_1$. This clearly shows a nonlinear relation between the velocity and spectral parameter. Therefore, we introduce the parameters $\xi_j$ to simplify our soliton solution, which is also much more convenient for discussing the soliton's physical properties. The white line corresponds to $w_1\neq0$ or $b_1\neq0$.}\label{FigA1}
\end{figure}
Under this representation, one can guarantee that the solitons' velocities in the first-step iteration \eqref{eq:first transformation} and the second-step iteration \eqref{eq:second transformation} remain exactly equal, i.e., $\widetilde{v}_2=\widetilde{v}_1$. In other words, by performing the iterations in \eqref{eq:first transformation} and \eqref{eq:second transformation} and setting the spectral parameters under the conditions \eqref{spectrum} and \eqref{constrain2}, the first component of $Q[2]$ is the DVDS solution.

As we have mentioned above, the velocity expression  plays a crucial role in the derivation process of multivalley dark solitons. However, for the spectral parameter with the form $\lambda_j=a_j+\textrm{i} b_j$, we must take the square root of a complex number to obtain the physical parameters, namely, the velocity and width. To facilitate this analysis, we employ a more direct approach. Interestingly, we note that if we describe the spectral parameter with the form $\lambda_j=\frac{1}{2}(\xi_j+\frac{1}{\xi_j})$ with an arbitrary complex parameter $\xi_j=-v_j+\textrm{i} w_j$, the solution can be simplified greatly. Moreover, the physical meanings of the parameters $v_j$ and $w_j$ are much clearer than those of the parameters $a_j$ and $b_j$, as shown in Fig.~\ref{FigA1}. Then the eigenvalues of Eq.\eqref{EV} can be re-expressed as
\begin{small}
\begin{eqnarray}
&&\tau_{j1}\!=-\!\frac{1}{2}\left[\xi_j\!-\!\frac{1}{\xi_j}\right], \tau_{j2}\!=\!-\frac{1}{2}\left[\xi_j\!+\!\frac{1}{\xi_j}\right],\nonumber\\
&&\tau_{j3}\!=\!-\frac{1}{2}\left[\xi_j\!+\!\frac{1}{\xi_j}\right],
\tau_{j4}\!=\!\frac{1}{2}\left[\frac{1}{\xi_j}\!-\!\xi_j\right].\nonumber
\end{eqnarray}
\end{small}
To simplify the solution forms, we can set the coefficients $\phi_{j1}$, $\phi_{j2}$, and $\phi_{j3}$ to
\begin{eqnarray}
&& c_{j1}=1,\\
&& c_{j2}=\frac{\xi_j+1}{\xi_j}\sqrt{1-|\xi_j|^2}\alpha_j,\\
&& c_{j3}=\frac{\xi_j+1}{\xi_j}\sqrt{1-|\xi_j|^2}\beta_j,
 \end{eqnarray}
where $\alpha_j$ and $\beta_j$ are arbitrary constants.
Then, the specific solution \eqref{eq:eigenfunctions2} of spectrum problem \eqref{Laxpair} at $\lambda_j$ can be re-expressed as the following form:
\begin{equation}\label{eq:eigenfunctions3}
\psi_j=\left(
         \begin{array}{c}
           \psi_{j1} \\
           \psi_{j2} \\
           \psi_{j3} \\
           \psi_{j4} \\
         \end{array}
       \right)=
\left(\!
  \begin{array}{c}
    (1+\xi_j) \phi_{j1}\\
    (1+\frac{1}{\xi_j})\phi_{j1}e^{-it} \\
    \phi_{j2} \\
    \phi_{j3} \\
  \end{array}
\!\right).
\end{equation}

Then, we can obtain the DVDS solution in the first component of $Q[2]$ and the bright solitons in the other two components by performing the following two-fold DT processes. For the first-step DT \eqref{eq:first transformation}, the spectral parameters are $\lambda_1=\frac{1}{2}(\xi_1+1/\xi_1)$ and $\xi_1=-v_1+\textrm{i}w_1$, and the eigenfunction is constrained by $\psi_{13}=\phi_{12}=0$, i.e., $\alpha_1=0$ (or  the eigenfunction $\psi_{14}=\phi_{13}=0$, i.e., $\beta_1=0$). For the second-step DT \eqref{eq:second transformation}, the spectral parameters are $\lambda_2=\frac{1}{2}(\xi_2+1/\xi_2)$ and $\xi_2=-v_1+\textrm{i}w_2$, and the eigenfunction is constrained by $\psi_{24}=\phi_{23}=0$, i.e., $\beta_2=0$  (or the eigenfunction $\psi_{23}=\phi_{23}=0$, i.e., $\alpha_2=0$). The simplified solution for $Q[2]$ \eqref{eq:second transformation} is presented in Eq. \eqref{DBB}, where $\beta_1=0,\alpha_2=0$, $\alpha_1$ and $\beta_2$ are nonzero constants. This indicates that the parameter $v_1$ is the velocity, and $w_1$ and $w_2$ are the width-dependent parameters for DVDSs. Therefore, the physical meanings of the parameters $v_j$ and $w_j$ are much clearer than those of the parameters $a_j$ and $b_j$ \eqref{spectrum}, as shown in Fig.~\ref{FigA1}.

To study the collision dynamics of DVDSs, further iterations are needed. For example, we can investigate the collision between a DVDS and an SVDS by performing a third-step DT with $\lambda_3=\frac{1}{2}(\xi_3+1/\xi_3)$ and $\xi_3=-v_2+\textrm{i}w_3$. We employ $\psi_3$, which is mapped to $\psi_3[2]=(T[2]\psi_3[1])|_{\lambda=\lambda_3}$ with $\psi_3[1]=(T[1]\psi_3)|_{\lambda=\lambda_3}$:
\begin{eqnarray}\label{eq:third transformation}
&&\psi[3]=T[3]\psi[2],T[3]=\mathbb{I}\!+\!\frac{\lambda_3^*\!-\!\lambda_3}{\lambda\!-\!\lambda_3^*}
    \frac{\psi_3[2]\psi_3[2]^\dag\Lambda}{\psi_3[2]^\dag\Lambda\psi_3[2]}, \nonumber\\
&& Q[3]=Q[2]+(\lambda_3-\lambda_3^*)\left[\sigma_3,\frac{\psi_3[2]\psi_3[2]^\dag\Lambda}
     {\psi_3[2]^\dag\Lambda\psi_3[2]}\right].
\end{eqnarray}
For this case, we constrain the eigenfunctions $\psi_{33}\neq0$ and $\psi_{34}\neq0$, i.e., $\alpha_3\neq0$ and $\beta_3\neq0$, and the eigenfunctions $\psi_{1}$ and $\psi_{2}$ are the same as in the first-step and second-step DT. A typical example of this case is shown in Figs.~\ref{Fig4}(a1)-(a3), demonstrating a striking collision process for which a DVDS is transformed into a breather after colliding with an SVDS.

Naturally, by performing a fourth-step DT, one can investigate the interaction between two DVDSs. Before performing this iteration, the eigenfunctions of $\psi_{3}$ in the third-step DT \eqref{eq:third transformation} should be made to satisfy the constraint conditions where the eigenfunction $\psi_{34}=\phi_{33}=0$, i.e., $\beta_3=0$ (or  the eigenfunction $\psi_{33}=\phi_{32}=0$, i.e., $\alpha_3=0$). Then, we employ $\psi_4$, which is mapped to $\psi_4[3]=(T[3]\psi_4[2])|_{\lambda=\lambda_4}=(T[3]T[2]T[1]\psi_4)|_{\lambda=\lambda_4}$. Then, the two DVDS solutions can be obtained as follows with the spectral parameters $\lambda_4=\frac{1}{2}(\xi_4+1/\xi_4)$ and $ \xi_4=-v_2+\textrm{i}w_4$, and we can constrain the eigenfunction $\psi_{43}=\phi_{42}=0$, i.e., $\alpha_4=0$ (or the eigenfunction $\psi_{44}=\phi_{43}=0$, i.e., $\beta_4=0$):
\begin{eqnarray}\label{eq:fourth transformation}
Q[4]=Q[3]+(\lambda_4-\lambda_4^*)\left[\sigma_3,\frac{\psi_4[3]\psi_4[3]^\dag\Lambda}
     {\psi_4[3]^\dag\Lambda\psi_4[3]}\right].\\\nonumber
\end{eqnarray}

In the four-component case, the TVDS can also be obtained by performing the above developed DT with three-fold iterations . We do not show the detailed derivation for this process herein. The simplified solution for TVDS is presented in \eqref{DBBB}.

In general, the $n$-fold Darboux matrix can be constructed in the following form:
\begin{equation}\label{eq:n-fold-dt}
\begin{split}
\mathbf{T}_n=\mathbb{I}+\mathbf{Y}_n\mathbf{M}_n^{-1}(\lambda\mathbb{I}-\mathbf{D}_n)^{-1}\mathbf{Y}_n^{\dag}\Lambda,
\end{split}
\end{equation}
with
\begin{align*}
\mathbf{Y}_n&=\left[|\psi_1\rangle,|\psi_2\rangle,\cdots, |\psi_n\rangle\right]=\left[\begin{array}{c}
                              \Psi_1 \\
                              \Psi_2
                            \end{array}
\right],\\
\mathbf{D}_n&={\rm diag}\left(\lambda_1^*,\lambda_2^*,\cdots,\lambda_n^*\right), \\
\mathbf{M}_n&=\left(\frac{\langle\psi_i|\psi_j\rangle}{\lambda_i^*-\lambda_j}\right)_{1\leq i,j\leq n}.
\end{align*}
$\Psi_1$ is a $1\times n$ matrix, and $\Psi_2$ is an $N\times n$ matrix.
The B\"acklund transformation between the old potential functions and the new functions is expressed as follows:
\begin{equation}\label{eq:back-n-fold}
\begin{split}
\mathbf{q}[n]&=\mathbf{q}+2\Psi_2 \mathbf{M}^{-1}\Psi_1^\dag.\\
\end{split}
\end{equation}
\section{}
The explicit expressions for $f_j$, $g_j$, $h_j$, and $m_j$ of the asymptotic expressions in Eqs.~\eqref{DBBb} are
\begin{small}
\begin{equation*}
\begin{aligned}
f_1&=\frac{\xi_1^*\xi_2^*\xi_3^*(w_1-w_2)^2}{\xi_1\xi_2\xi_3(w_1+w_2)^2}\sigma_5\sigma_6,f_2=\frac{\xi_3^*}{\xi_3}\sigma_7\sigma_8\alpha_1^2\beta_2^2,\\
f_3&=\frac{\xi_1^*\xi_3^*}{\xi_1\xi_3}\beta_2^2\sigma_5\sigma_8,f_4=\frac{\xi_2^*\xi_3^*}{\xi_2\xi_3}\alpha_1^2\sigma_6\sigma_7,\\
g_1&=\frac{2\alpha_1 w_1p_1}{v_1-\textrm{i}w_1}\left[2(v_1-v_2)w_3+\textrm{i}\sigma_1\right],\\
g_2&=\frac{w_1-w_2}{w_1+w_2}\sigma_2,g_3=\sigma_3\beta^2,\\
h_1&=\frac{2\beta_2 w_2 p_2}{v_1-\textrm{i}w_2}\left[2(v_2-v_1)w_3-\textrm{i}\sigma_2\right],\\
h_2&=\frac{w_1-w_2}{w_1+w_2}\sigma_1,h_3=\sigma_4\alpha^2,\\
m_1&=\frac{(w_1-w_2)^2}{(w_1+w_2)^2}\sigma_5\sigma_6,m_2=\sigma_7\sigma_8\alpha_1^2\beta_2^2,\\
m_3&=\sigma_6\sigma_7\alpha_1^2,m_4=\sigma_5\sigma_8\beta_2^2,\\
p_1&=\sqrt{1-|\xi_1|^2},p_2=\sqrt{1-|\xi_2|^2},\\
\xi_1&=-v_1+\textrm{i}w_1,\xi_2=-v_1+\textrm{i}w_2,\xi_3=-v_2+\textrm{i}w_3,\\
\sigma_1&=(v_1\!-\!v_2)^2\!+\!(w_1^2\!-\!w_3^2),\sigma_2\!=\!(v_1\!-\!v_2)^2\!+\!(w_2^2\!-\!w_3^2),\\
\sigma_3&=(v_1\!-\!v_2)^2\!+\!(w_2^2\!+\!w_3^2),\sigma_4\!=\!(v_1\!-\!v_2)^2\!+\!(w_1^2\!+\!w_3^2),\\
\sigma_5&=(v_1\!-\!v_2)^2\!+\!(w_1\!-\!w_3)^2,\sigma_6\!=\!(v_1\!-\!v_2)^2\!+\!(w_2\!-\!w_3)^2,\\
\sigma_7&=(v_1\!-\!v_2)^2\!+\!(w_1\!+\!w_3)^2,\sigma_8\!=\!(v_1\!-\!v_2)^2\!+\!(w_2\!+\!w_3)^2.
\end{aligned}
\end{equation*}
\end{small}

The explicit expressions for $\delta_j$, $\zeta_j$, $\varsigma_j$, and $\varrho_j$ of the asymptotic expressions in Eqs.~\eqref{DBBa} are
\begin{small}
\begin{equation*}
\begin{aligned}
R_j&\!=\!|1+\xi_j|^2,S_j\!=\!1+\xi_j^2,\Lambda_{ij}=\xi_i\xi_j^*-1,\Xi_{ij}=\xi_i-\xi_j^*,\\
\Gamma_j&\!=\!(\xi_j+\xi_j^*)(1+|\xi_j|^2),G_j=\Xi_{ij}\Lambda_{ij},(i,j=1,2,3)\\
\end{aligned}
\end{equation*}
\end{small}
\begin{small}
\begin{equation*}
\begin{aligned}
\delta_1&\!=\!p_1p_2\alpha_1\beta_2\frac{R_1R_2|\Lambda_{21}|^2G_{11}G_{22}G_{33}}{8|\xi_1|^2|\xi_2|^2\xi_1\xi_2},\\
\delta_2&\!=\!\frac{\Xi_{21}}{|\xi_1|^2G_{31}G_{23}},\delta_3=\frac{\Xi_{12}}{|\xi_2|^2G_{32}G_{13}},\\
\delta_4&\!=\!\frac{R_1R_2|\Lambda_{21}|^2\Lambda_{11}\Lambda_{22}}{4|\xi_1|^2|\xi_2|^2\xi_1\xi_2},\delta_5=\frac{|\xi_1-\xi_2|^2}{\xi_1\xi_2},\\
\delta_6&\!=\!\frac{-\alpha_1^2\beta_2^2|\Xi_{12}|^2}{4\xi_1^*\xi_2^*|G_{13}|^2|G_{32}|^2}\big[G_{11}G_{22}G_{33}^2-(2|\xi_3|^2\!|S_1|^2\!+\\
&\!2|\xi_1|^2\!|S_3|^2\!-\!\Gamma_1\!\Gamma_3)(2|\xi_3|^2\!|S_2|^2\!+\!2|\xi_2|^2\!|S_3|^2\!-\!\Gamma_2\!\Gamma_3\!)\big]\!,\\
\delta_7&\!=\!\frac{R_1R_2\Lambda_{11}\Lambda_{22}|G_{12}|^2}{8|\xi_1|^2|\xi_2|^2\xi_1\xi_2},\\
\delta_8&\!=\!\alpha_1^2\frac{2|\xi_3|^2|S_1|^2+2|\xi_1|^2|S_3|^2-\Gamma_1\Gamma_3}{\xi_1^*\xi_2|G_{13}|^2},\\
\delta_9\!&=\!\beta_2^2\frac{2|\xi_3|^2|S_2|^2+2|\xi_2|^2|S_3|^2-\Gamma_2\Gamma_3}{\xi_1\xi_2^*|G_{23}|^2},\\
\zeta_1\!&=\!\frac{p_1\alpha_1 R_1R_2G_{12}\Lambda_{22}}{8\xi_1\xi_2^*\xi_3|\xi_2|^2G_{13}},\\
\zeta_2\!&=\!\frac{\xi_1^*-\xi_2^*}{|\xi_1|^4\xi_2}G_{11}\Lambda_{21}\left[\xi_1\Gamma_3-2S_1|\xi_3|^2\right],\\
\zeta_3\!&=\!\frac{8\beta_2^2}{|G_{23}|^2}(\lambda_1-\lambda_1^*)(\lambda_1^*-\lambda_2)\big[\big(4\lambda_1|\xi_3|^2-\Gamma_3\big)\\
&\!\left(2|S_2|^2|\xi_3|^2+2|S_3|^2|\xi_2|^2-\Gamma_2\Gamma_3\right)\!+\!G_{22}G_{33}^2\big],\\
\zeta_4\!&=\!\frac{p_2\beta_2 R_1R_2\Xi_{21}|\Lambda_{21}|^2\Lambda_{11}G_{22}G_{33}}{8|\xi_1|^4|\xi_2|^4\xi_3G_{23}},\zeta_5\!=\!\xi_2^*-\xi_1^*,\\
\zeta_6\!&=\!\alpha_1^2\Xi_{12}[|\xi_3|^2\xi_1^*S_1\Lambda_{11}+|\xi_1|^2S_1\Gamma_3-\xi_1|\xi_1|^2R_3\\
&-|\xi_1|^2|\xi_3|^2S_1(\xi_1+\xi_1^*)]/[\xi_1|G_{13}|^2],\\
\varsigma_1\!&=\!p_1\alpha_1\frac{ R_1R_2G_{12}G_{33}\Lambda_{22}}{8\xi_1\xi_2^*\xi_3|\xi_2|^2G_{13}},\varsigma_2\!=\!\frac{\xi_1^*-\xi_2^*}{|\xi_1|^2\xi_1^*\xi_2}G_{11}\Lambda_{21},\\
\varsigma_3\!&=\!\frac{16\beta_2^2}{|G_{23}|^2}(\!\lambda_1^*\!-\!\lambda_1\!)\!(\!\lambda_1^*\!-\!\lambda_2\!)\!\big[S_2^2|\xi_3|^2\xi_2^*/\xi_2\!-\!\xi_2^*S_2\Gamma_3\!+\!|\xi_2|^2R_3\big],\\
\varsigma_4\!&=\! \frac{-p_2\beta_2(\xi_2^*-\xi_2)}{8|\xi_1|^2|\xi_2|^2\xi_2\xi_3G_{23}}R_1R_2|\Lambda_{21}|^2G_{22}\Lambda_{11}\Xi_{21},\\
\varsigma_5\!&=\!(\xi_2^*-\xi_1^*)(2S_2|\xi_3|^2-\xi_2\Gamma_3),\\
\varsigma_6\!&=\!\alpha_1^2\Xi_{12}\big[\!-\xi_2G_{11}G_{33}^2-(2S_2|\xi_3|^2-\xi_2\Gamma_3)\\
&(2|\xi_3|^2R_1+2|\xi_1|^2R_3-\Gamma_1\Gamma_3)\big]/(2|G_{13}|^2),\\
\varrho_1\!&=\!p_1p_2\alpha_1\beta_2\frac{R_1R_2G_{11}G_{22}|\Lambda_{21}|^2}{8|\xi_1|^4|\xi_2|^4|G_{13}^2G_{32}|^2},\\
\varrho_2\!&=\!G_{32}G_{13}G_{33}\Xi_{23},\\
\varrho_3\!&=\!\frac{R_1R_2\Lambda_{11}\Lambda_{22}|\Lambda_{21}|^2}{4|\xi_1|^4|\xi_2|^4},\varrho_4\!=\!|\xi_1-\xi_2|^2,\\
\varrho_5\!&=\!\frac{-\alpha_1^2\beta_2^2|\Xi_{12}|^2}{4|G_{13}|^2|G_{23}|^2}\big[-(2|\xi_3|^2R_1+2|\xi_1|^2R_3-\Gamma_1\Gamma_3)\\
&(2|\xi_3|^2R_2+2|\xi_2|^2R_3-\Gamma_2\Gamma_3)+G_{11}G_{22}G_{33}^2\big],\\
\varrho_6\!&=\!\beta_2^2|\Xi_{12}|^2\big[2|\xi_3|^2R_2+2|\xi_2|^2R_3-\Gamma_2\Gamma_3\big]/(2|G_{23}|^2),\\
\varrho_7\!&=\!\alpha_1^2|\Xi_{12}|^2\big[2|\xi_3|^2R_1+2|\xi_1|^2R_3-\Gamma_1\Gamma_3\big]/(2|G_{13}|^2).
\end{aligned}
\end{equation*}
\end{small}

\end{appendix}

\end{document}